\documentclass[journal]{IEEEtran}
\usepackage[utf8]{inputenc}
\usepackage[T1]{fontenc}

\usepackage[pdftex]{graphicx}
\usepackage{pgfplots}
\pgfplotsset{compat=newest}%
\usepackage{tikz}
\usetikzlibrary{plotmarks,external,positioning}
\usetikzlibrary{external,shapes,chains,scopes,decorations.pathreplacing,calc,backgrounds}%
\makeatletter
\def\ps@IEEEtitlepagestyle{
	\def\@oddfoot{\mycopyrightnotice}
	\def\@evenfoot{}
}
\def\mycopyrightnotice{
	{\footnotesize
		\begin{minipage}{\textwidth}
			\centering
			\textcopyright~2018 IEEE. Personal use is permitted, but republication/redistribution requires IEEE permission. \\
			See \url{http://www.ieee.org/publications_standards/publications/rights/index.html} for more information.
		\end{minipage}
	}
}

\makeatother

\usepackage{amsmath,mathrsfs,amssymb,amsthm,bm,mathtools,pifont,siunitx}
\newtheorem{Lemma}{Lemma}

\newcounter{MYtempeqncnt}
\newcounter{MYtempeqncnt2}

\ifCLASSOPTIONcompsoc
\usepackage[caption=false,font=normalsize,labelfont=sf,textfont=sf]{subfig}
\else
\usepackage[caption=false,font=footnotesize]{subfig}
\fi
\usepackage[bookmarks=false]{hyperref}
\usepackage{cite}

\usepackage{ifthen}
\usepackage{booktabs}
\usepackage{csquotes}

\newcommand{\expectation}[2][]{%
	\ifthenelse{\equal{#1}{}}{\mathrm{E}\!\left[#2\right]}{\mathrm{E}_{#1}\!\!\left[#2\right]}%
}
\newcommand{\tr}[1]{\operatorname{tr}\left(#1\right)}
\newcommand{\trsq}[1]{\operatorname{tr}^2\left(#1\right)}

\newcommand{\diag}[2]{\operatorname{diag}_{#1}\!{(#2)}}
\newcommand{\Shat}{\hat{\mathbf{S}}}
\newcommand{\id}{\,\mathrm{d}}
\newcommand{\btilde}[1]{\tilde{\mathbf{#1}}}
\newcommand{\bhat}[1]{\hat{\mathbf{#1}}}
\renewcommand{\S}{\bm{\Sigma}}
\newcommand{\C}{\bm{\Gamma}}

\newlength\smallfigurewidth
\newlength\smallfigureheight
\newlength\matrixsize
\usepackage{dblfloatfix} 

\begin{document}

\title{LMPIT-inspired Tests for Detecting a Cyclostationary Signal in Noise with Spatio-Temporal Structure}

\author{Aaron~Pries,~\IEEEmembership{Student~Member,~IEEE,} %
	David~Ramírez,~\IEEEmembership{Senior~Member,~IEEE,} %
	and~Peter~J.~Schreier,~\IEEEmembership{Senior~Member,~IEEE}%
	\thanks{This research was supported by the German Research Foundation (DFG) under grant SCHR 1384/6-1. The work of D. Ram{\'\i}rez has been partly supported by Ministerio de Econom{\'\i}a of Spain under projects: OTOSIS (TEC2013-41718-R) and the COMONSENS Network (TEC2015-69648-REDC), by the Ministerio de Econom{\'\i}a of Spain jointly with the European Commission (ERDF) under projects ADVENTURE (TEC2015-69868-C2-1-R) and CAIMAN (TEC2017-86921-C2-2-R), by the Comunidad de Madrid under project CASI-CAM-CM (S2013/ICE-2845), and by the German Research Foundation (DFG) under project RA 2662/2-1.}%
	\thanks{A. Pries and P. J. Schreier are with the Signal \& System Theory Group, Paderborn University, Paderborn, Germany (e-mail: aaron.pries@sst.upb.de; peter.schreier@sst.upb.de).}%
	\thanks{D. Ramírez is with the Signal Processing Group, Universidad Carlos III de Madrid, Leganés, Spain, and the Gregorio Marañón Health Research Institute, Madrid, Spain (e-mail: david.ramirez@uc3m.es).}%
}

\maketitle%
\begin{abstract}
	In spectrum sensing for cognitive radio, the presence of a primary user can be detected by making use of the cyclostationarity property of digital communication signals. For the general scenario of a cyclostationary signal in temporally colored and spatially correlated noise, it has previously been shown that an asymptotic generalized likelihood ratio test (GLRT) and locally most powerful invariant test (LMPIT) exist. In this paper, we derive detectors for the presence of a cyclostationary signal in various scenarios with structured noise. In particular, we consider noise that is temporally white and/or spatially uncorrelated. Detectors that make use of this additional information about the noise process have enhanced performance. We have previously derived GLRTs for these specific scenarios; here, we examine the existence of LMPITs. We show that these exist only for detecting the presence of a cyclostationary signal in spatially uncorrelated noise. For white noise, an LMPIT does not exist. Instead, we propose tests that approximate the LMPIT, and they are shown to perform well in simulations. Finally, if the noise structure is not known in advance, we also present hypothesis tests using our framework.
\end{abstract}

\begin{IEEEkeywords}
Cyclostationarity, detection, generalized likelihood ratio test (GLRT), interweave cognitive radio, locally most powerful invariant test (LMPIT), spectrum sensing
\end{IEEEkeywords}

\ifCLASSOPTIONdraftcls
\fi
\section{Introduction}
\IEEEPARstart{D}{etection} of cyclostationarity has received renewed attention in recent years. A particularly interesting application is interweave cognitive radio \cite{Axell2012}. This technology contributes to a more efficient use of the electromagnetic spectrum by sensing wireless channels, such that unlicensed secondary users can opportunistically access radio resources. The signal transmitted by the primary user is \emph{unknown} to the secondary user, but nevertheless the detection has to perform reliably even for low SNR. 

For the detection of a primary user in noise, there exist many models and corresponding tests (e.g.\ \cite{Yucek2009,Ariananda2009,Lim2008,Zeng2009,Font-Segura2010,Sharma2015,Ali2017}). Existing detectors include energy detectors (e.g.\ \cite{Yucek2009,Ariananda2009,Sharma2015,Ali2017}), eigenvalue detectors (e.g.\ \cite{Zeng2009,Sharma2015}), correlation-based detectors (e.g.\ \cite{Yucek2009,Lim2008,Sharma2015,Ali2017}) and others. Some detectors are based on the generalized likelihood ratio test (GLRT), and many assume white noise. Which detector is applicable for a particular scenario depends on the information available about the primary-user signal. A more general scenario was considered in \cite{Ramirez2011}, where the noise is allowed to be spatially uncorrelated and temporally colored. These papers, however, do not exploit the prior information that the signal of interest is a digital communication signal, which is \emph{cyclostationary} (e.g.\ \cite{Gardner1987a}), while the noise is wide-sense stationary (WSS). This enables us to build better detectors by detecting this cyclostationarity feature. If it cannot be found, we conclude that only noise is present. For an introduction to cyclostationarity in general, and its detection in particular, the reader is referred to the  papers \cite{Gardner1994,Spooner1994,Gardner2006,Napolitano2016,Axell2012}. %

Early detectors for cyclostationarity were developed in \cite{Gardner1992, Dandawate1994, Spooner1994, Enserink1995}, and since cognitive radio has become a popular idea, more detectors have been proposed, e.g.\ \cite{Lunden2009, Huang2013, Urriza2013}. Recent publications have also proposed detectors for particular classes of primary-user signals, for example BPSK \cite{Ly2017}, OFDM \cite{Kosmowski2017}, and GFDM \cite{El-Alfi2017}. Our goal in this paper is to develop detectors of cyclostationarity for arbitrary modulation schemes. A problem related to the \emph{detection} of signals is the \emph{identification} of the modulation scheme, where it also is possible to utilize the cyclostationarity feature of communications signals \cite{Spooner2000, Dobre2007, Dobre2012, Marey2012}. \par
The classical detector for the presence of cyclostationarity is given in \cite{Dandawate1994} and similar detectors for observations from multiple antennas are proposed in \cite{Lunden2009,Huang2013}. These detectors test whether cycles are present in the autocorrelation function for a specified set of lags. The tests from \cite{Spooner1994,Enserink1995,Urriza2013,Sedighi2014} use the fact that spectral components of a cyclostationary process are correlated for some lags/frequencies. The choice of these lags is commonly optimized in advance, but this may not be possible in a cognitive radio framework, where we do not have prior information about the signal of the primary user. 

A different family of detectors, where only the cycle period needs to be known, was derived in \cite{Ramirez2014a,Ramirez2015b}. These detectors can inherently deal with observations from multiple antennas, but they are based on the assumption of having available independent complex normally-distributed observations. 
This assumption may sound restrictive at first, but it enables the use of powerful statistical methods. While normality is necessary to derive the detectors, the covariance matrices need not be known. In practice, independently distributed observations can be approximately obtained by chopping one long observation into multiple short observations.
The detectors in \cite{Ramirez2014a} and \cite{Ramirez2015b} are an asymptotic GLRT and an asymptotic locally most powerful invariant test (LMPIT) for a low-SNR scenario. Interestingly, both tests are different functions of the same coherence matrix. Further, both proposed detectors outperform classical detectors even when applied to the detection of communication signals, which are not Gaussian.

While these detectors assume arbitrarily colored and spatially correlated noise under the null hypothesis, the noise might have further structure in the context of cognitive radio. In a properly calibrated system, noise is temporally white and spatially uncorrelated, which is commonly assumed. Yet calibration may also fail in either one of these domains, leading to noise that is only temporally white or spatially uncorrelated. In this paper, we treat all possible cases, i.e., temporally white and/or spatially uncorrelated noise. These scenarios were already considered in \cite{Pries2016}, which derived GLRTs for each of them. It turns out that the GLRTs are again a function of a coherence matrix, but the coherence matrix is defined differently from \cite{Ramirez2015b} in order to account for the additional structure of the noise. Other tests for the detection of a cyclostationary process in white noise were developed by \cite{Axell2011,Riba2014}. For the case of white noise, the proposed GLRT of \cite{Pries2016} results in a substantially improved performance compared to either the GLRTs in \cite{Ramirez2015b} or the general-noise detectors in \cite{Urriza2013,Lunden2009}. Our paper investigates the existence of (locally) optimal tests for the same assumptions about the noise as in \cite{Pries2016}, i.e.\ for temporally white and/or spatially uncorrelated noise.\par%
\subsection{Contributions}
The main contributions of this paper can be summarized as follows:
\begin{enumerate}
	\item We propose detectors for an arbitrary cyclostationary signal with known cycle period in noise that is temporally white and/or spatially uncorrelated. By incorporating the additional information about the noise into our model, we are able to derive detectors with improved performance. Our detectors do not require knowledge of the signal parameters.
	\item We investigate whether locally (i.e. low-SNR) \emph{optimal} tests, that is, LMPITs, exist for these scenarios. For temporally colored but spatially uncorrelated noise the LMPIT exists, and we derive its closed-form expression. For temporally white noise, such a test does not exist as it depends on unknown quantities. Instead we propose LMPIT-inspired tests. For all tests we derive an approximate distribution under the null hypothesis, which allows us to choose the thresholds of the tests.
	\item We give an interpretation of our LMPIT-inspired tests in terms of the cyclic spectrum. We show that our detectors use the cyclic coherence function, generalized to multiple antennas/time series, and show how they utilize different cyclic frequencies.
	\item We validate our detectors in simulations using different setups for the signal and noise. We show that the LMPIT for spatially uncorrelated noise outperforms both the GLRT and the LMPIT for correlated noise. For temporally white noise, we demonstrate that our proposed tests also outperform other state-of-the-art detectors. Furthermore we evaluate the computational complexity of our detectors.
\end{enumerate}

\subsection{Outline}
Our program for this paper is the following: In \autoref{sec:problem}, we formulate the problem and an asymptotic approximation thereof. Then we derive the structure of the hypothesis test for the various noise assumptions. We review the GLRTs for these problems in \autoref{sec:GLRT}. In \autoref{sec:LMPIT}, we analyze the existence of LMPITs for the different scenarios. Based on these results, we propose LMPIT-inspired detectors for the case of temporally white noise (\autoref{sec:LMPIT_inspired}). All proposed tests are evaluated and compared to state-of-the-art detectors by numerical simulations in \autoref{sec:simulations}. We derive the computational complexity of our detectors in \autoref{sec:complexity}. Finally, in \autoref{sec:characterization}, we propose tests to determine the spatial and temporal structure of the noise, which can be used to select the appropriate test for cyclostationarity.
\section{Problem Formulation}
\label{sec:problem}
We consider the detection of a discrete-time cyclostationary signal with known cycle period $P\in\mathbb{N}\setminus\{1\}$ in the presence of noise with spatio-temporal structure.\footnote{If the cycle period is not known in advance, it can be estimated, for example with the estimators in \cite{Dandawate1994,Ramirez2014}. Making our detectors robust against cycle-period mismatch is beyond the scope of this paper but it could possibly be achieved following along the lines of \cite{Horstmann2017}. Other existing robust solutions can be found in \cite{Zeng2010,Sharma2015a,Kumar2013}.} %
Denoting the observations from $L$ time series by the vector $\mathbf{x}[n]\in\mathbb{C}^L$, cyclostationarity means that the autocorrelation function is periodic in the global time variable $n$:
\begin{equation}
\expectation{\mathbf{x}[n]\mathbf{x}^H[n-k]} \coloneqq \mathbf{M}[n,k] = \mathbf{M}[n+P,k].
\end{equation}
We stack $NP$ consecutive samples of $\mathbf{x}[n]$ into the vector 
\begin{equation}
\mathbf{y} = \left[\mathbf{x}^T[0],\, \dots,\, \mathbf{x}^T[NP-1]\right]^T.
\end{equation}
Based on $\mathbf{y}$, the goal is to decide whether or not the observed process is cyclostationary. We assume that $\mathbf{x}[n]$, and thus $\mathbf{y}$, is a proper complex Gaussian random variable with zero mean. The covariance matrix ${\mathbf{R}}=\expectation{\mathbf{yy}^H}$ of $\mathbf{y}$ depends on the autocorrelation sequence $\mathbf{M}[n,k]$:
\begin{equation}
	{\mathbf{R}} = \begin{bmatrix}{{\mathbf{M}}[0,0]} & \dots& {{\mathbf{M}}[0,-NP+1]}\\
	\vdots & \ddots & \vdots \\
	{{\mathbf{M}}[NP-1,NP-1]} & \dots& {{\mathbf{M}}[NP-1,0]}\end{bmatrix}.
\end{equation}
Any additional information about ${\mathbf{M}}[n,k]$ will result in a particular structure of the covariance matrix $\mathbf{R}$. 

Regarding the observation $\mathbf{x}[n]$, we have two scenarios: Either a signal is present and then $\mathbf{x}[n]$ is cyclostationary, or only noise is observed and then $\mathbf{x}[n]$ is WSS. This is described by the hypotheses
\begin{equation}
\begin{split}
\mathcal{H}_1&: \text{$\mathbf{x}[n]$ is cyclostationary,} \\
\mathcal{H}_0&: \text{$\mathbf{x}[n]$ is WSS, and additionally temporally white}\\
&\quad\text{and/or spatially uncorrelated.}
\end{split}
\label{eq:firstHypothesis}
\end{equation}
\enquote{Spatial} correlation has to be interpreted as the correlation between the different time series in the vector $\mathbf{x}[n]$. Because of the Gaussian assumption, we use the following equivalent formulation of \eqref{eq:firstHypothesis} in terms of the covariance matrix of $\mathbf{y}$:
\begin{equation}
 \begin{split}
 \mathcal{H}_1&: \mathbf{y} \sim \mathcal{CN}(\mathbf{0},\mathbf{R}_1) \\
 \mathcal{H}_0&: \mathbf{y} \sim \mathcal{CN}(\mathbf{0},\mathbf{R}_0),
 \end{split}
 \label{eq:secondHypothesis}
\end{equation}
where $\mathcal{CN}$ denotes the proper complex Gaussian distribution. Thus the hypotheses only differ in the covariance matrix, and we are therefore interested in the structure of $\mathbf{R}_1$ and $\mathbf{R}_0$. To determine this structure, we define the autocorrelation functions of $\mathbf{x}[n]$ for the respective hypotheses as
\begin{subequations}
\begin{align}
\mathbf{M}_1[n,k] &\coloneqq \expectation[\mathcal{H}_1]{\mathbf{x}[n]\mathbf{x}^H[n-k]}\\
\mathbf{M}_0[k] &\coloneqq \expectation[\mathcal{H}_0]{\mathbf{x}[n]\mathbf{x}^H[n-k]}.
\end{align}
\label{eq:correlations}
\end{subequations}
Under $\mathcal{H}_1$, the cyclostationarity of the signal means that $\mathbf{M}_1[n,k]$ is periodic in $n$:
\begin{equation}
\mathbf{M}_1[n,k] = \mathbf{M}_1[n+P,k],
\end{equation}
where $P$ is the cycle period. This periodicity causes the covariance matrix $\mathbf{R}_1$ to be block-Toeplitz with block-size $LP$ \cite{Gladyshev1961,Ramirez2015b}. %

Under the null hypothesis, the considered spatio-temporal information about the noise process results in distinct properties of $\mathbf{M}_0[k]$: In the case of spatially uncorrelated noise, $\mathbf{M}_0[k]$ is diagonal for all $k$. White noise results in an $\mathbf{M}_0[k]$ that is zero except for the lag $k=0$. If the noise is temporally white and spatially uncorrelated, we further know that $\mathbf{M}_0[0]$ is diagonal. In terms of the covariance matrix $\mathbf{R}_0$, all considered scenarios of the noise result in a block-Toeplitz $\mathbf{R}_0$, with block-size $L$ \cite{Pries2016}:~In the case of spatially uncorrelated noise, all $L\times L$ \emph{blocks} will be \emph{diagonal}. For temporally white noise, the matrix $\mathbf{R}_0$ becomes \emph{block-diagonal} and the combination of white and uncorrelated noise will cause the whole matrix to be \emph{diagonal}. %

We could now test between the two hypotheses based on the structure of the covariance matrix, if the correlation functions \eqref{eq:correlations}, or equivalently, $\mathbf{R}_1$ and $\mathbf{R}_0$, were completely known. However, since we do not have any knowledge about them (besides the structure), this is a \emph{composite} hypothesis test. Common approaches for this type of test are the GLRT, the uniformly most powerful invariant test (UMPIT), or the LMPIT. For the particular case of the GLRT, this poses a problem because closed-form maximum likelihood (ML) estimates of block-Toeplitz matrices do not exist \cite{Burg1982}. Therefore, we follow the approach from \cite{Ramirez2015b}, where we approximate the block-Toeplitz matrices $\mathbf{R}_0$ and $\mathbf{R}_1$ as block-circulant matrices, denoted by $\mathbf{Q}_0$ and $\mathbf{Q}_1$. This means that $\mathbf{Q}_0$ and $\mathbf{Q}_1$ can be block-diagonalized by DFT matrices, and thus be estimated in closed form. For this, the vector $\mathbf{y}$ is transformed into the frequency domain using
\begin{equation}
\mathbf{z} = \left(\mathbf{L}_{NP,N}\otimes\mathbf{I}_L\right)\left(\mathbf{F}_{NP}\otimes\mathbf{I}_L\right)^H\mathbf{y},
\label{eq:transformation}
\end{equation}
where $\mathbf{L}_{NP,N}$ is the commutation matrix, defined such that $\operatorname{vec}\left(\mathbf{A}\right) = \mathbf{L}_{NP,N}\operatorname{vec}\left(\mathbf{A}^T\right)$ for a $P\times N$ matrix $\mathbf{A}$ \cite{Magnus2007}. Further, $\mathbf{F}_{NP}$ is the $NP$-dimensional DFT-matrix, and $\otimes$ denotes the Kronecker product. The linear transformation \eqref{eq:transformation} then block-diagonalizes $\mathbf{Q}_0$ and $\mathbf{Q}_1$, and we can express the hypotheses as
\begin{equation}
\begin{split}
\mathcal{H}_1&: \mathbf{z} \sim \mathcal{CN}(\mathbf{0},\mathbf{S}_1) \\
\mathcal{H}_0&: \mathbf{z} \sim \mathcal{CN}(\mathbf{0},\mathbf{S}_0).
\end{split}
\label{eq:thirdHypothesis}
\end{equation}
	\begin{figure*}
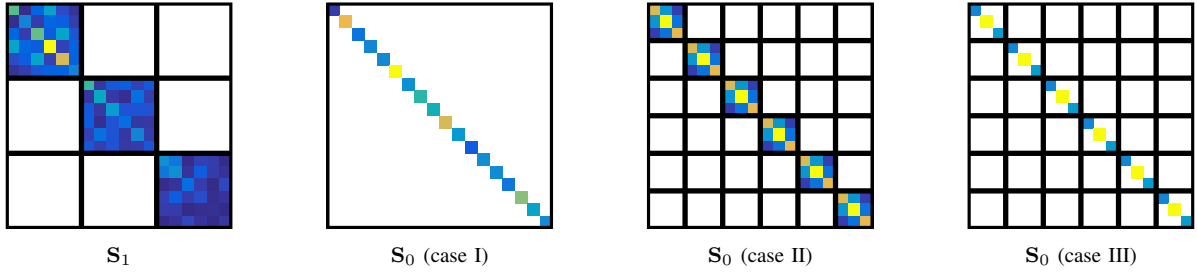

		\centering
		\tikzsetnextfilename{structure}
		\begin{tikzpicture}
		\node (A) {\input{S_1}};
		\node[node distance=0.0cm,below=of A] (A_text) {\footnotesize$\mathbf{S}_1$};
		\node[right=of A] (B) {\input{S_0_uncorrelated}};
		\node[node distance=0.0cm,below=of B] (B_text) {\footnotesize$\mathbf{S}_0$ (case I)};
		
		\node[right=of B] (C) {\input{S_0_white}};
		\node[node distance=0.0cm,below=of C] (C_text) {\footnotesize$\mathbf{S}_0$ (case II)};
		\node[right=of C] (D) {\input{S_0_both}};
		\node[node distance=0.0cm,below=of D] (D_text) {\footnotesize$\mathbf{S}_0$ (case III)};
\end{tikzpicture}%
		\caption{Structure of the covariance matrices, with $L = 3$, $P= 2$, and $N = 3$. Case I is temporally colored and spatially uncorrelated noise, case II is white and correlated noise, and case III is white and uncorrelated noise.}%
		\label{fig:structure}
	\end{figure*}
The transformation \eqref{eq:transformation} is designed such that the covariance matrix of $\mathbf{z}$ becomes asymptotically (for $N\rightarrow\infty$) block-diagonal under both hypotheses. The covariance matrix $\mathbf{S}_1$ has diagonal blocks of size $LP\times LP$, and $\mathbf{S}_0$ is block-diagonal with blocks of size $L\times L$ \cite{Ramirez2015b}. As before, we obtain further structure under $\mathcal{H}_0$ depending on the assumption about the noise: For the case of spatially uncorrelated and temporally colored noise, the whole matrix $\mathbf{S}_0$ becomes diagonal (case I).  For white and spatially correlated noise (case II), the diagonal blocks are identical and thus $\mathbf{S}_0$ can be factorized as $\mathbf{I}_{NP}\otimes\tilde{\mathbf{S}}_0$, where the matrix $\tilde{\mathbf{S}}_0$ is unknown. In the case of temporally white and spatially uncorrelated noise, these blocks are also diagonal (case III) \cite{Pries2016}. The structure of the covariance matrix $\mathbf{S}_0$ for all considered cases is illustrated in Figure~\ref{fig:structure}.
\subsection{Comparison with related problems}
For temporally colored but spatially uncorrelated noise, the hypotheses in \eqref{eq:thirdHypothesis} differ only in the block-size of the covariance matrices. An LMPIT for hypotheses with such a structure was already derived in \cite{Ramirez2015b}, so the results can be immediately applied to this problem. More details will be presented in \autoref{sec:uncorrelated}.

For the case of temporally white noise, the covariance matrix is block-diagonal under $\mathcal{H}_1$ and $\mathcal{H}_0$, and the block-sizes are $LP$ and $L$, respectively. Under $\mathcal{H}_0$, however, the blocks are identical. Apart from this structure, the only additional information we have is that all blocks are positive definite. In a related paper for the detection of cyclostationarity in WSS noise \cite{Ramirez2015b}, the structure is very similar, but under $\mathcal{H}_0$, the blocks are \emph{not} identical.

At the same time, the structure of the white-noise scenario is related to another problem, where the covariance matrix is positive definite under $\mathcal{H}_1$ and block-diagonal with the same positive definite blocks under $\mathcal{H}_0$. This scenario was considered in \cite{Ramirez2013}, and the present problem is a generalization thereof. For the special case of $N=1$, the two problems are identical. However, for $N > 1$ this problem is much more difficult and, as we will show, the LMPIT does not exist.

\section{GLRTs for Structured Noise}%
\label{sec:GLRT}
In this section, we review the GLRTs for detecting a cyclostationary signal in WSS noise with further spatio-temporal structure. We originally derived these in \cite{Pries2016}, and here we also present a way to set the threshold of the tests to achieve a particular probability of false alarm. \par%
To apply the GLRT to observed data, we assume to have  $M\geq LP$ independent and identically distributed (i.i.d.) realizations $\mathbf{z}_i$ of the vector $\mathbf{z}$. In practice, often there is only one observation available. In such a case, we would split the whole observation (assumed to be of length $MNP$) into $M$ signals of length $NP$. Formally, this violates the assumption of independence, but as we will show in later simulations, this does not affect the performance much. Splitting the whole observation into segments can be interpreted in light of Bartlett's method of estimating the (cyclic) power spectral density\cite{Bartlett1948,Bartlett1950}. It sacrifices resolution in the frequency domain in order to decrease the variance of the estimators. The ratio between $M$ and $N$ thus controls the tradeoff between these two effects. \par
\begin{table}[t]
	\centering
	\caption{Estimate of $\mathbf{S}_0$ for different noise assumptions}
	\label{tab:S0_hat}
	\begin{tabular}{@{} *3c @{}}
		\toprule
		\emph{noise structure}  & $\bhat{S}_0$   & \\ \midrule
		colored \& correlated & $\diag{L}{\bhat{S}}$                                                 \\
		\begin{tabular}{c}colored \& uncorrelated \\ (case I)\end{tabular}& $\diag{}{\bhat{S}}$                                                                 \\
		\begin{tabular}{c}white \& correlated   \\ (case II)\end{tabular}   & $\mathbf{I}_{NP}\otimes\frac{1}{NP}\sum\limits_{k=0}^{NP-1}\bhat{S}^{(k,k)}$          \\
		\begin{tabular}{c}white \& uncorrelated  \\ (case III)\end{tabular}  & $\mathbf{I}_{NP}\otimes\frac{1}{NP}\sum\limits_{k=0}^{NP-1}\diag{}{\bhat{S}^{(k,k)}}$ \\ \bottomrule\hline
	\end{tabular}
\end{table}
For the generalized likelihood ratio, which is defined as
\begin{equation}
	\mathscr{L}_G = \frac{\underset{\mathbf{S}_0}{\max}\enspace p(\mathbf{z}_1,\hdots,\mathbf{z}_M;\mathbf{S}_0)}{\underset{\mathbf{S}_1}{\max}\enspace p(\mathbf{z}_1,\hdots,\mathbf{z}_M;\mathbf{S}_1)},
	\label{eq:GLR}
\end{equation}%
we need the ML estimates of the unknown covariance matrices. They depend on the sample covariance matrix 
\begin{equation}
\bhat{S} = \frac{1}{M}\sum\limits_{i=0}^{M-1}\mathbf{z}_i\mathbf{z}_i^H.
\label{eq:S_full}
\end{equation}
The ML-estimate of $\mathbf{S}_1$ is \cite{Ramirez2015b}
\begin{align}
	\bhat{S}_1 &= \diag{LP}{\bhat{S}},
\end{align}
where the $\diag{B}{\cdot}$ operator returns the diagonal blocks of size $B$ and sets the off-diagonal blocks to zero. For the case of $B=1$, we will use $\diag{}{\cdot}$. Under $\mathcal{H}_0$, the likelihood function can be written as
\begin{align}
	&\pi^{-LMNP}\prod_{k=0}^{NP-1}\left(\det\mathbf{S}_0^{(k,k)}\right)^{-M}\nonumber\\
	&\qquad\times\exp\left\{-M\tr{\sum_{k=0}^{NP-1}\left(\mathbf{S}_0^{(k,k)}\right)^{-1}\bhat{S}^{(k,k)}}\right\}
\end{align}
for all cases, where $(\cdot)^{(k,k)}$ denotes the $(k,k)$th $L\times L$ block of a matrix. For matrix blocks and elements, we use indexing starting from zero. Depending on the structure of the noise, the blocks $\mathbf{S}_0^{(k,k)}$ have further structure, as outlined in \autoref{sec:problem}. This leads to the ML estimates of $\mathbf{S}_0$ as derived in \cite{Pries2016} and listed in \autoref{tab:S0_hat}. The case of temporally colored and spatially correlated noise was covered in \cite{Ramirez2015b} and is listed for the sake of completeness.\par%
With these estimates plugged into \eqref{eq:GLR}, the GLRTs for the different scenarios can all be expressed as
\begin{equation}
	\mathscr{L}_G  \propto \det(\bhat{C}) \underset{\mathcal{H}_1}{\overset{\mathcal{H}_0}{\gtrless}} \eta,
	\label{eq:GLRT}
\end{equation}
with the sample coherence matrix 
\begin{equation}
	\bhat{C} =  \bhat{S}_0^{-1/2}\bhat{S}_1\bhat{S}_0^{-1/2}.
	\label{eq:C_hat}
\end{equation}
For the interpretation of the blocks of $\bhat{C}$ in terms of the cyclic power spectral density (PSD), see the remarks in \cite[Section VI]{Ramirez2015b}. The threshold $\eta$ can be obtained for a given false alarm rate using Wilks' theorem \cite{Wilks1938,Ramirez2015b}: According to Wilks, $-2M\log\det(\bhat{C})$ is asymptotically $\chi^2$-distributed under the null hypothesis, with degrees of freedom depending on the number of parameters to be estimated under the two hypotheses. For the cases considered in this paper, the degrees of freedom are listed in \autoref{tab:DOF}.
\begin{table}[!t]
	\centering
	\caption{Degrees of freedom of the $\chi^2$-distribution.}
	\label{tab:DOF}
	\begin{tabular}{@{} *3c @{}}
		\toprule
		\emph{noise structure}  & degrees of freedom   & \\ \midrule
		colored \& correlated & $L^2NP(P-1)$ \\
		\begin{tabular}{c}colored \& uncorrelated \\ (case I)\end{tabular}&             $LNP(LP-1)$                                               \\
		\begin{tabular}{c}white \& correlated   \\ (case II)\end{tabular}   &          $L^2(NP^2-1)$ \\
		\begin{tabular}{c}white \& uncorrelated  \\ (case III)\end{tabular}  & $L(LNP^2-1)$\\ \bottomrule\hline
	\end{tabular}
\end{table}
\section{LMPITs for Structured Noise}
\label{sec:LMPIT}
As in \cite{Ramirez2013,Ramirez2015b}, we use Wijsman's theorem \cite{Wijsman1967} to find an expression for the LMPIT. With this theorem, it is possible to express the likelihood ratio of the maximal invariant statistic without the distribution of the likelihood of the maximal invariant. If the resulting expression only depends on known quantities or observations, we obtain a UMPIT \cite{Scharf1990}. If this is not the case, we can seek approximations in order to find an LMPIT, which is only locally optimal. Such an LMPIT for the case of a multivariate cyclostationary process in WSS noise with arbitrary spatio-temporal correlations was derived in \cite{Ramirez2015b}. In the following subsections, we discuss the case where more specific information about the noise is available.
%
\ifCLASSOPTIONdraftcls
\else
\begin{figure*}[!t]%
	\hrulefill
	\setcounter{MYtempeqncnt}{\value{equation}}
	\setcounter{equation}{19} 
	\input{eq_wijsman.tex}
	\setcounter{equation}{\value{MYtempeqncnt}}
	\hrulefill
\end{figure*}
\fi %
\subsection{LMPIT for Temporally Colored and Spatially Uncorrelated Noise}
\label{sec:uncorrelated}
In the case of temporally colored and spatially uncorrelated noise, we test between two block-diagonal covariance matrices that differ only in their block-size. In \cite{Ramirez2015b}, the LMPIT was derived for two arbitrary block sizes and here it is applied to our problem. Hence the statistic of the LMPIT is
\begin{equation}
\mathscr{L}_u = \left\lVert
\left(\diag{}{\bhat{S}}\right)^{-1/2}\bhat{S}_1\left(\diag{}{\bhat{S}}\right)^{-1/2}
\right\rVert^2_F\,,
\label{eq:LMPIT_uncorrelated}
\end{equation}
where $\left\lVert\cdot\right\rVert_F$ denotes the Frobenius norm of a matrix. The LMPIT is obtained by comparing $\mathscr{L}_u$ with a threshold $\eta_u$:
\begin{equation}
\mathscr{L}_u \underset{\mathcal{H}_0}{\overset{\mathcal{H}_1}{\gtrless}} \eta_u.
\end{equation}
As for the tests in \cite{Ramirez2015b}, $M(\mathscr{L}_u-LNP)$ is approximately $\chi^2$-distributed under the null hypothesis, with the same degrees of freedom as the GLRT in \autoref{tab:DOF}, i.e. $LNP(LP-1)$.
\subsection{LMPIT for Temporally White and Spatially Correlated Noise}
\label{sec:LMPIT_white}
If the noise is assumed to be temporally white and spatially correlated, the covariance matrix under the null hypothesis is block-diagonal with identical blocks of size $L$. As before, the structure of the covariance matrix under the alternative hypothesis is block-diagonal with block-size $LP$. To find an optimal invariant test for this scenario using Wijsman's theorem \cite{Wijsman1967}, we first need to identify the problem invariances as a group. For the structure of the covariance matrix under the hypotheses as listed in Section \ref{sec:problem}, the group $\mathcal{G}$ is
\begin{equation}
\mathcal{G} = \{\mathbf{z}\rightarrow g(\mathbf{z}): g(\mathbf{z})= \left(\mathbf{P}\otimes\mathbf{Q}\otimes\mathbf{G}\right)\mathbf{z}\},
\label{eq:inv_group}
\end{equation}
where $\mathbf{P}$ is an $N\times N$ permutation matrix, $\mathbf{Q}$ is a $P\times P$ unitary matrix, and $\mathbf{G}$ is an $L\times L$ nonsingular matrix. To keep the notation concise, we define $\tilde{\mathbf{G}} = \mathbf{P}\otimes\mathbf{Q}\otimes\mathbf{G}$, and the sets of permutation, unitary, and nonsingular matrices are denoted by $\mathbb{P}$, $\mathbb{Q}$, and $\mathbb{G}$, respectively. 

A transformation from this group leaves the structure of the hypotheses unchanged and can be interpreted as follows: 
Since the covariance matrix is block-diagonal with unknown $LP\times LP$ blocks under both hypotheses (see Section~\ref{sec:problem} and Figure~\ref{fig:structure}), the blocks on the diagonal can be permuted arbitrarily without changing the block-diagonal structure. This is captured by the matrix $\mathbf{P}$ and represents a frequency reordering. To see the effect of $\left(\mathbf{Q}\otimes\mathbf{G}\right)$, we have to look at the structure of the diagonal $LP\times LP$ blocks of $\mathbf{S}$ under the two hypotheses, i.e.\ either $\mathbf{S}_0$ or $\mathbf{S}_1$. If we denote their $j$th block by $\mathbf{S}_j$, then the group action transforms this block to
\begin{equation}
\left(\mathbf{Q}\otimes\mathbf{G}\right)\mathbf{S}_j\left(\mathbf{Q}^H\otimes\mathbf{G}^H\right).
\label{eq:transf_block}
\end{equation}
Under $\mathcal{H}_1$, $\mathbf{S}_j$ is an unknown and unstructured matrix, and this is not affected by the group action. Under $\mathcal{H}_0$, the block $\mathbf{S}_j$ is itself a block-diagonal matrix, with identical blocks on the diagonal, i.e.\ it can be written as $\mathbf{I}_P\otimes\mathbf{A}$. Then, according to \eqref{eq:transf_block}, the transformed block becomes $\mathbf{I}_P\otimes\mathbf{GAG}^H$, which is still block-diagonal with identical blocks (transformed by $\mathbf{G}$) on the diagonal. \par
Applying Wijsman's theorem, we obtain %
\ifCLASSOPTIONdraftcls%
\begin{align}
	\displaystyle\mathscr{L} &= \frac
	{\displaystyle\sum\limits_{\mathbb{P}}\int\limits_{\mathbb{Q}}\int\limits_{\mathbb{G}}\det(\mathbf{S}_1)^{-M}\lvert\det\mathbf{G}\rvert^{2MNP}\operatorname{exp}\left\{-M\tr{\mathbf{S}_1^{-1}\tilde{\mathbf{G}}\hat{\mathbf{S}}\tilde{\mathbf{G}}^H}\right\}\id\mathbf{G}\id\mathbf{Q}}
	{\displaystyle\sum\limits_{\mathbb{P}}\int\limits_{\mathbb{Q}}\int\limits_{\mathbb{G}}\det(\mathbf{S}_0)^{-M}\lvert\det\mathbf{G}\rvert^{2MNP}\operatorname{exp}\left\{-M\tr{\mathbf{S}_0^{-1}\tilde{\mathbf{G}}\hat{\mathbf{S}}\tilde{\mathbf{G}}^H}\right\}\id\mathbf{G}\id\mathbf{Q}}
	\label{eq:wijsman}
\end{align}%
\else%
\stepcounter{equation}%
(\ref{eq:wijsman}) on the top of the next page 
\fi%
for the ratio of the distributions of the maximal invariant statistic. This expression is now simplified and similar to the GLRT in \eqref{eq:GLRT}, Equation~\eqref{eq:wijsman} can be written as a function of the sample coherence matrix $\bhat{C}$:
\begin{Lemma}
	\label{lemma_wijsman}
	The ratio of the distributions of the maximal invariant statistic \eqref{eq:wijsman} can be written as
	\begin{align}
	\mathscr{L} &\propto \sum\limits_{\mathbb{P}}\int\limits_{\mathbb{Q}}\int\limits_{\mathbb{G}}\beta\left(\mathbf{G}\right)\operatorname{e}^{-\alpha}\id\mathbf{G}\id\mathbf{Q}\,,
	\label{eq:lemma_wijsman_start}
	\end{align}
	with $\alpha$ defined as
	\begin{align}
	\alpha &= M\tr{\mathbf{W}\hat{\mathbf{C}}},
	\end{align}
	which is a function of the observations and the matrices forming the group $\mathcal{G}$ as follows:
	\begin{align}
	\bhat{C} &=  \bhat{S}_0^{-1/2}\bhat{S}_1\bhat{S}_0^{-1/2},\\%
	\bhat{S}_0 &= \mathbf{I}_{NP}\otimes\frac{1}{NP}\sum\limits_{j=0}^{N-1}\sum\limits_{k=0}^{P-1}\bhat{S}_j^{(k,k)}, \\%
	\mathbf{W} &= \tilde{\mathbf{G}}^H(\tilde{\mathbf{S}}_1-\mathbf{I})\tilde{\mathbf{G}}\,, \\
	\tilde{\mathbf{S}}_1 &= (\mathbf{I}_{NP}\otimes\bar{\mathbf{S}}_1^{-1/2})\mathbf{S}_1^{-1}(\mathbf{I}_{NP}\otimes\bar{\mathbf{S}}_1^{-1/2})\,,\\
	\bar{\mathbf{S}}_1&= \frac{1}{NP}\sum\limits_{j=0}^{N-1}\sum_{k=0}^{P-1}\mathbf{S}_{1,j}^{(k,k)}\,,\\
	\beta\left(\mathbf{G}\right) &= \left\vert\det\mathbf{G}\right\vert^{2MNP}\exp\left\{-MNP\tr{\mathbf{GG}^H}\right\}.
	\end{align}
	\begin{proof}
		Please refer to Appendix~\ref{ap:lemma1}.
	\end{proof}
	\label{lemma1}
\end{Lemma}

Since the expression in \eqref{eq:lemma_wijsman_start} depends on the unknown parameters in $\btilde{S}_1$, a UMPIT does not exist. We can, however, approximate the exponential term for a low-SNR scenario \cite{Ramirez2015a} and check if the integral depends on unknowns. If it does not, then we have found an LMPIT. For low SNR (or more general: close hypotheses), we obtain $\tilde{\mathbf{S}}_1 \approx \mathbf{I}$ and thus $\alpha\approx 0$. This approximation is used to perform a Taylor series expansion of $\operatorname{exp}(-\alpha)$ around $\alpha = 0$:
\begin{equation}
\operatorname{exp}(-\alpha) \approx 1 - \alpha + \frac{1}{2} \alpha^2.
\end{equation}
By continuing with this approximation, we can no longer obtain a globally optimal test. All the remaining results will hold only approximately for a low SNR condition, which is particularly interesting for a cognitive radio application. Plugging this approximation into (\ref{eq:lemma_wijsman_start}), we obtain a sum of three terms, where the constant term can be discarded as it does not depend on the data. The remaining linear and quadratic terms will be dealt with in the following two lemmas.
\begin{Lemma}
	\label{lemma_linear}
	The linear term 
	\begin{equation}
	\sum\limits_{\mathbb{P}}\int\limits_{\mathbb{Q}}\int\limits_{\mathbb{G}}\beta(\mathbf{G})\tr{\mathbf{W}\hat{\mathbf{C}}}\id\mathbf{G}\id\mathbf{Q}
	\label{eq:LMPIT_approx_linear}
	\end{equation}
	in the Taylor series expansion of \eqref{eq:lemma_wijsman_start} is constant with respect to observations.
	\begin{proof}
		Please refer to Appendix~\ref{ap:lemma2}.
	\end{proof}
\end{Lemma}
Since the linear term does not depend on data, we can neglect it for the expression of the LMPIT. Consequently, we simplify the approximation of \eqref{eq:lemma_wijsman_start} by keeping only the quadratic term, i.e.
\begin{align}
\mathscr{L} \propto \sum\limits_{\mathbb{P}}\int\limits_{\mathbb{Q}}\int\limits_{\mathbb{G}}\beta(\mathbf{G})\trsq{\mathbf{W}\hat{\mathbf{C}}}\id\mathbf{G}\id\mathbf{Q}.
\label{eq:quadratic_term}
\end{align}
This term can be expressed in terms of the diagonal blocks $\bhat{C}_j$ of $\bhat{C}$ as stated in the following lemma:
\begin{Lemma}
	\label{lemma_quadratic}
	For $N>1$, the quadratic term in \eqref{eq:quadratic_term}  in the Taylor series expansion of \eqref{eq:lemma_wijsman_start} can be written as
	\begin{align}
	\mathscr{L} &\propto \sum\limits_{j=0}^{N-1}\|\hat{\mathbf{C}}_j\|_F^2 + \lambda P\sum\limits_{j=0}^{N-1}\|\hat{\bar{\mathbf{C}}}_j\|_F^2+\mu N\|\hat{\mathbf{C}}_{\mathrm{av}}\|_F^2,
	\label{eq:LMPIT}
	\end{align}
	with
	\begin{align}
	\hat{\bar{\mathbf{C}}}_j &= \frac{1}{P}\sum\limits_{k=0}^{P-1}\bhat{C}_j^{(k,k)},\\
	\hat{\mathbf{C}}_{\mathrm{av}} &= \frac{1}{N}\sum\limits_{j=0}^{N-1}\bhat{C}_j,\label{eq:C_av}
	\end{align}
	where $\bhat{C}_j^{(k,k)}$ denotes the $k$th $L\times L$ sub-block of the $j$th block $\bhat{C}_j$. The scalar quantities $\lambda$ and $\mu$ are constant with respect to observations, but they depend on unknown quantities in $\mathbf{W}$ and $\btilde{S}_1$, respectively.
	\begin{proof}
		Please refer to Appendix~\ref{ap:lemma3}.
	\end{proof}
\end{Lemma}
Since this expression still depends on unknown quantities, the LMPIT does not exist. 

\subsection{LMPIT for Temporally White and Spatially Uncorrelated Noise}
First we observe that the structure of $\mathbf{S}_0$ is very similar for temporally white noise that is either spatially correlated or uncorrelated. In both cases, the matrix is block-diagonal with repeating blocks. While the blocks are just positive definite matrices for correlated noise, these blocks become \emph{diagonal} with positive diagonal elements for uncorrelated noise. If the noise is assumed spatially uncorrelated, this constrains $\mathbf{G}$ from the group of invariances in \eqref{eq:inv_group} to become \emph{diagonal} with nonzero diagonal elements. The derivation of the ratio of the distributions of the maximal invariant statistic follows as in Section~\ref{sec:LMPIT_white}, while considering the additional constraint. In the end, the test statistic can be written as in \eqref{eq:LMPIT} if we replace
\begin{equation}
\bhat{S}_0 \rightarrow \diag{}{\bhat{S}_0}
\end{equation}
and use \eqref{eq:C_hat} to obtain the coherence matrix. For the same reason mentioned in the previous section, an LMPIT does not exist for this scenario, either.

\section{LMPIT-inspired Tests for White Noise}
\label{sec:LMPIT_inspired}
For a theoretical analysis of the test statistic \eqref{eq:LMPIT}, we need its probability density function (pdf). Since it seems very difficult to derive the pdf, we perform Monte Carlo simulations, in order to analyze how the test statistic \eqref{eq:LMPIT} performs if we use a grid of values for the unknown quantities $\lambda$ and $\mu$. This can be done only in simulations where we know whether a signal is present or not. Testing multiple values of the parameters cannot be done in a real-world application. However, this kind of simulation can reveal which of the terms contribute most (on average) towards a good detector. %
In particular, we answer two questions: How do the detectors based on the individual terms in \eqref{eq:LMPIT} perform? How do these tests perform compared to the white-noise GLRT \eqref{eq:GLRT} and the test based on \eqref{eq:LMPIT} with optimized values of $\lambda$ and $\mu$? The performance of a test will be measured by the area under the receiver operating characteristic (ROC) curve. \par

For all simulations, the observations are generated as 
\begin{equation}
	\begin{split}
		\mathcal{H}_1&: \mathbf{x}[n] = (\mathbf{H}*\mathbf{s})[n]+\mathbf{w}[n]\\
		\mathcal{H}_0&: \mathbf{x}[n] = \mathbf{w}[n],
	\end{split}
\end{equation}
where $\mathbf{s}[n]$ is a baseband QPSK signal with rectangular pulse shaping. A new symbol is drawn every $T$ samples, which causes the cycle period of $\mathbf{s}[n]$ to be $P=T$. The operation $(\mathbf{H}*\mathbf{s})[n]$ denotes a convolution with the channel $\mathbf{H}[n]$, which is a Rayleigh fading channel with exponential power delay profile, uncorrelated among antennas, and constant for each Monte Carlo experiment. However, a new realization of $\mathbf{H}[n]$ is drawn for each simulation and thus we average over many (on the order of $10\text{k}$) channels $\mathbf{H}[n]$. Finally, $\mathbf{w}[n]$ is temporally white Gaussian noise with spatial correlation.

\begin{figure}[!tb]
	\centering
	\ifCLASSOPTIONdraftcls
	\includegraphics[width = 0.7\linewidth]{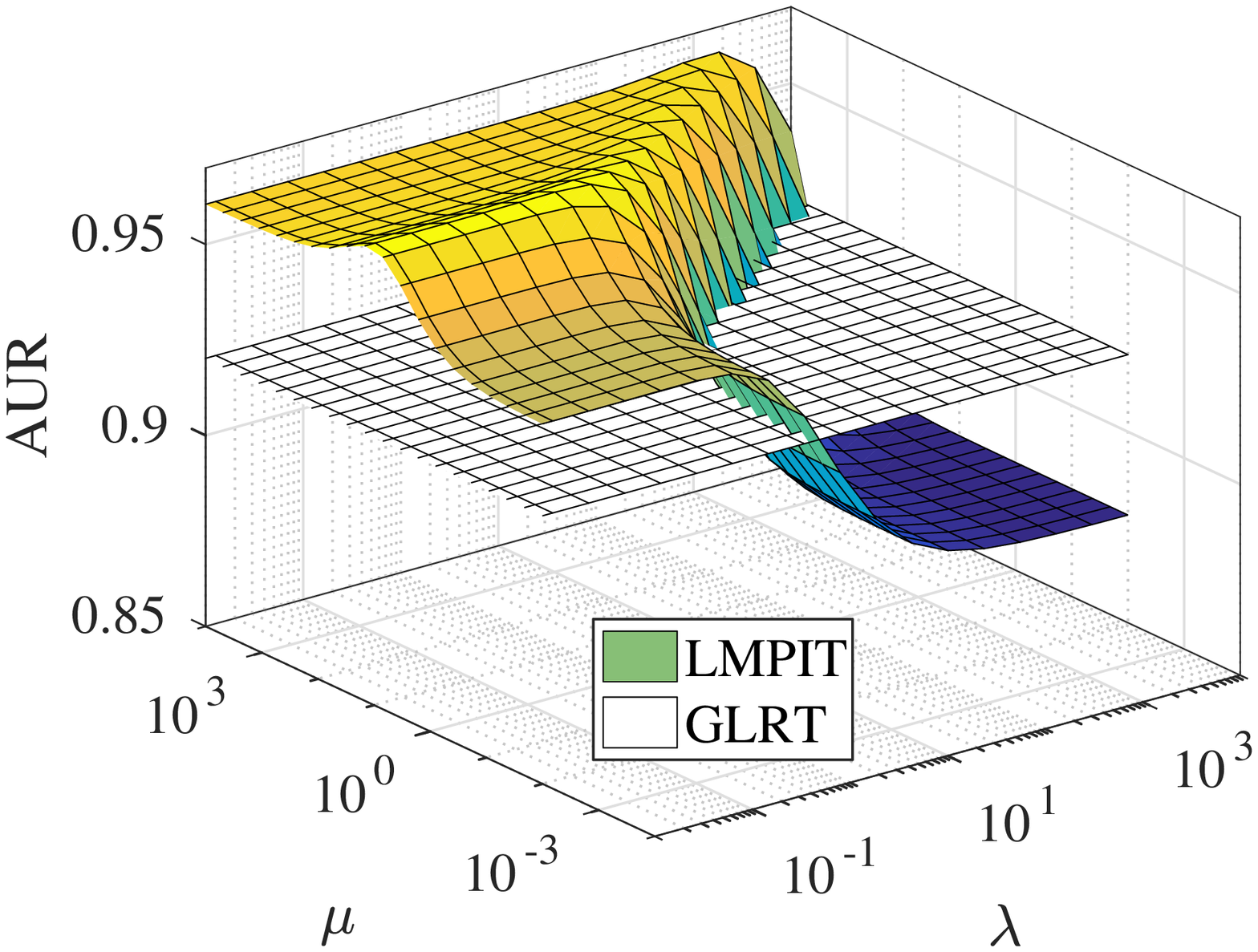}
	\else
	\includegraphics[width = 0.95\linewidth]{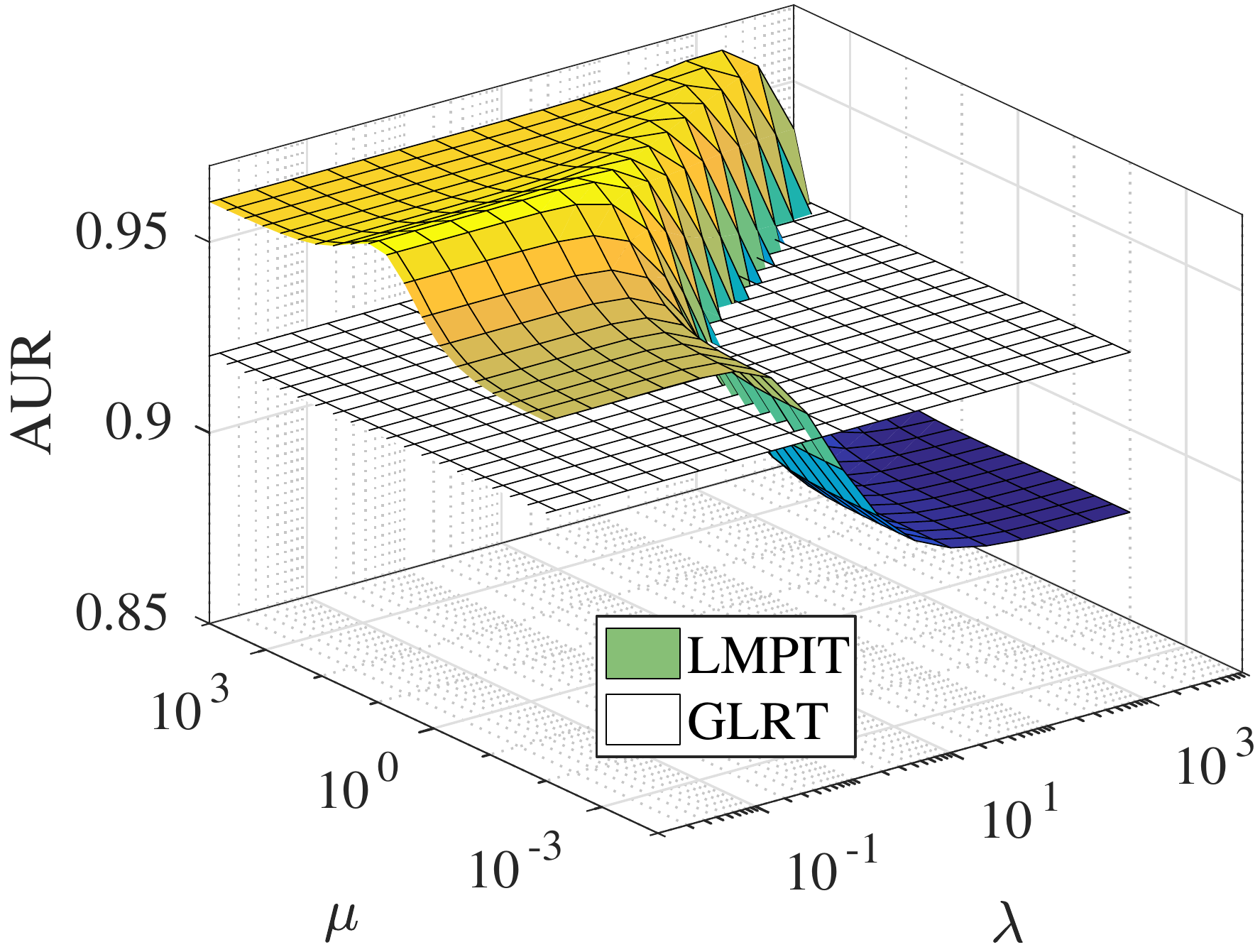}
	\fi
	\caption{Area under the ROC (AUR) of the test based on \eqref{eq:LMPIT} as a function of $\lambda$ and $\mu$. Performance of the GLRT \eqref{eq:GLRT} for reference. The parameters $N=64$, $\mathrm{SNR}=\SI{-15}{dB}$, $P=3$, $L=3$, and $M=20$ are used.}
	\label{fig:AUR}
\end{figure}
An example is seen in \autoref{fig:AUR}, where the LMPIT-curve reveals that the optimal values for $\mu$ and $\lambda$ for this particular scenario are close to 12 and 0.6, respectively. For different simulation setups these values vary, but extensive experiments have shown that they are never very large nor close to zero. 
If the maximum were in a corner, which corresponds to either a large or a small parameter, we would obtain the best test by using only one of the terms in \eqref{eq:LMPIT}. These individual terms can be found in \autoref{fig:AUR}: Using large values of $\lambda$ and small values of $\mu$ is equivalent to approximating the test statistic by the term $\sum_{j=0}^{N-1}\|\hat{\bar{\mathbf{C}}}_j\|_F^2$ alone, which performs worse than the GLRT. Using small $\lambda$ and either small $\mu$ or large $\mu$ is approximately the same as using either the term $\sum_{j=0}^{N-1}\|\hat{\mathbf{C}}_j\|_F^2$ or $\|\hat{\mathbf{C}}_{\mathrm{av}}\|_F^2$, respectively, as test statistic without the need to choose particular values of $\lambda$ and $\mu$ (which cannot be determined in general). Since the tests based on these statistics perform better than the GLRT, we propose them as \emph{LMPIT-inspired} tests for the case of temporally white noise.

These LMPIT-inspired tests have suboptimal performance. However, according to \autoref{fig:AUR} the performance can still be substantially better than competing tests, as the comparison with the GLRT \eqref{eq:GLRT} demonstrates. More detailed simulations, as well as comparisons with other detectors, will be presented in Section \ref{sec:simulations}.

The distribution of these statistics under the null hypothesis can be obtained by the relationship between the log-det and the Frobenius norm \cite{Leshem2001, Ramirez2015b}. Since the GLR for white noise in \autoref{sec:GLRT} was $\det(\bhat{C})^M=\prod_{j=0}^{N-1}\det(\hat{\mathbf{C}}_j)^M$, we can conclude that
\begin{equation}
M\left( \sum_{j=0}^{N-1}\|\hat{\mathbf{C}}_j\|_F^2 - LNP\right)
\end{equation}
is approximately $\chi^2$-distributed under the null hypothesis, with degrees of freedom as in \autoref{tab:DOF}.\par
Concerning the second proposed statistic, i.e.\ $\|\hat{\mathbf{C}}_{\mathrm{av}}\|_F^2$, it can be shown that $\det(\hat{\mathbf{C}}_\mathrm{av})^{MN}$ is the GLR for the hypotheses
\begin{equation}
	\begin{split}
	\mathcal{H}_1&: \mathbf{z} \sim \mathcal{CN}(\mathbf{0},\mathbf{I}_N\otimes\mathbf{S}_{LP}) \\
	\mathcal{H}_0&: \mathbf{z} \sim \mathcal{CN}(\mathbf{0},\mathbf{I}_{NP}\otimes\mathbf{S}_L)
	\end{split}
	\label{eq:hypotheticalHypothesis}
\end{equation}
where $\mathbf{S}_{LP}$ and $\mathbf{S}_{P}$ are unknown matrices of dimension $LP$ and $L$, respectively. %
Thus we can argue again that this log-GLR is asymptotically $\chi^2$-distributed, this time with $L^2(P^2-1)$ degrees of freedom. Thus for the case of temporally white and spatially correlated noise, the modified statistic
\begin{equation}
MN\left( \|\hat{\mathbf{C}}_{\mathrm{av}}\|_F^2 - LP\right)
\label{eq:mod_C_av}
\end{equation}
is approximately $\chi^2$-distributed with $L^2(P^2-1)$ degrees of freedom. The case of temporally white and spatially \emph{un}correlated noise differs only in the degrees of freedom, which are $L(LP^2-1)$.
\subsection{Interpretation of the LMPIT-inspired Tests}
The proposed LMPIT-inspired detectors can be interpreted in terms of the cyclic PSD of the random process $\mathbf{x}[n]$. The cyclic PSD can be written as \cite{Ramirez2015b,Schreier2010}
\begin{equation}
\S^{(c)}(\theta)d\theta = \expectation{d\bm{\xi}(\theta)d\bm{\xi}^H\left(\theta-\frac{2\pi c}{P}\right)},
\end{equation}
where $c/P$ denotes the cycle frequencies, and $d\bm{\xi}(\theta)$ is an increment of the random spectral process that generates $\mathbf{x}[n]$:
\begin{equation}
\mathbf{x}[n] = \int_{-\pi}^{\pi} d\bm{\xi}(\theta)e^{j\theta n}.
\end{equation}
As shown in \cite{Ramirez2015b,Schreier2010}, the covariance matrix $\mathbf{S} = \expectation{\mathbf{zz}^H}$ contains samples of the cyclic PSD $\S^{(c)}(\theta)$ at some frequencies $\theta$ and $c\in\mathbb{Z}$. 
Under the alternative hypothesis, $\mathbf{S}$ is block-diagonal with blocks of size $LP$. If we denote the $L\times L$ sub-blocks of the $j$th $LP\times LP$ block $\mathbf{S}_j$ by $\mathbf{S}_j^{(k,\kappa)}$, it turns out that 
\begin{equation}
\mathbf{S}_j^{(k,\kappa)} = \S^{(k-\kappa)}\left(\frac{2\pi(\kappa N+j)}{NP}\right)
\end{equation}
holds for $j=0,\dots,N-1$, $k=0,\dots,P-1$, and $\kappa=0,\dots,P-1$. Under the null hypothesis, a similar relation holds. In this case, $\mathbf{S}$ is block-diagonal with blocks of size $L$ and these blocks are identical because the noise is white:
\begin{equation}
\mathbf{S}_j^{(k,k)} = \S^{(0)}\left(\frac{2\pi(k N+j)}{NP}\right) = \S^{(0)}
\end{equation}
Thus the diagonal blocks only contain the cyclic PSD for the cycle frequency zero, which is the standard PSD.

After defining the coherence function
\begin{equation}
\C^{(c)}\left(\theta\right) = \left(\S^{(0)}\right)^{-1/2}
\S^{(c)}\left(\theta\right)
\left( \S^{(0)}\right)^{-1/2},
\end{equation}
the blocks of the coherence matrix $\mathbf{C}$ can be written as
\begin{align}
	\mathbf{C}_j^{(k,\kappa)} 
	&= \C^{(k-\kappa)}\left(\frac{2\pi(\kappa N+j)}{NP}\right).
\end{align}
The \emph{sample} coherence matrix $\bhat{C} = \bhat{S}_0^{-1/2}\bhat{S}_1\bhat{S}_0^{-1/2}$ consists of the blocks 
\begin{equation}
\bhat{C}_j^{(k,\kappa)} = \hat{\bar{\mathbf{S}}}_0^{-1/2}\bhat{S}_j^{(k,\kappa)}\hat{\bar{\mathbf{S}}}_0^{-1/2},
\end{equation}
where $\bhat{S}_j^{(k,\kappa)}$ can be interpreted as an estimate of the cyclic PSD:
\begin{equation}
\bhat{S}_j^{(k,\kappa)} =  \hat{\S}^{(k-\kappa)}\left(\frac{2\pi(\kappa N+j)}{NP}\right)
\end{equation}
Further, 
\begin{equation}
\hat{\bar{\mathbf{S}}}_0 = \frac{1}{NP}\sum_{j=0}^{N-1}\sum_{k=0}^{P-1}\bhat{S}_j^{(k,k)} = \hat{\S}^{(0)}
\end{equation}
is an estimate of the PSD in the case of white noise.

Now we can express the first proposed test statistic in terms of the sample coherence function $\hat{\C}^{(c)}\left(\theta\right)$:
\begin{align}
	\sum\limits_{j=0}^{N-1}\|\bhat{C}_j\|^2_F = 
	&\sum_{j=0}^{NP-1}\left\|\hat{\C}^{(0)}\left(\frac{2\pi j}{NP}\right)\right\|^2_F \nonumber
	\\
	&\quad+2\sum_{c=1}^{P-1}\sum_{j=0}^{(P-c)N-1}\left\|\hat{\C}^{(c)}\left(\frac{2\pi j}{NP}\right)\right\|^2_F
\end{align}
Hence this statistic accounts for both temporal correlation, as measured by the term with $c=0$, and the degree of cyclostationarity ($c\neq 0$). To find an interpretation for the second proposed test statistic, we define
\begin{equation}
\hat{\bar{\S}}^{(c)}_\kappa = \frac{1}{N}\sum_{j=0}^{N-1}\hat{\S}^{(c)}\left(\frac{2\pi(\kappa N+j)}{NP}\right),
\end{equation}
which is the block-wise average of the cyclic PSD. Then it turns out that the second proposed test statistic can be written as a function of $\hat{\bar{\C}}^{(c)}_\kappa =  \left(\hat{\S}^{(0)}\right)^{-1/2}\hat{\bar{\S}}^{(c)}_\kappa\left(\hat{\S}^{(0)}\right)^{-1/2}$:
\begin{align}
	\begin{split}
		\|\bhat{C}_\text{av}\|^2_F = 
		\sum_{\kappa=0}^{P-1}\left\|\hat{\bar{\C}}^{(c)}_\kappa\right\|^2_F
		+2\sum_{c=1}^{P-1}\sum_{\kappa=0}^{P-c-1}\left\|\hat{\bar{\C}}^{(c)}_\kappa\right\|^2_F
	\end{split}
\end{align}
This test statistic also measures the amount of color and the amount of cyclostationarity, but with an averaged coherence function.
\section{Simulations}
\label{sec:simulations}
In this section we compare the LMPIT for uncorrelated noise and the LMPIT-inspired tests for white noise with different detectors. Unless mentioned otherwise, the observations are generated as stated in \autoref{sec:LMPIT_inspired}.
\subsection{Spatially Uncorrelated Noise}
We first compare the performance of the GLRT \eqref{eq:GLRT}, the proposed LMPIT based on \eqref{eq:LMPIT_uncorrelated}, and the LMPIT from \cite{Ramirez2015b}, which is based on the more general assumption of correlated noise. The simulation setup is the same as introduced in \autoref{sec:LMPIT_inspired}, in this case using spatially uncorrelated and temporally colored noise $\mathbf{w}[n]$. Colored noise is realized by passing a temporally white signal through a moving average filter of the length $19$. The parameters are chosen as $\mathrm{SNR}=\SI{-17}{dB}$, $P = 3$, $N = 64$, $L = 3$, and $M = 10$. 

\begin{figure}[!t]
	\centering
		\tikzsetnextfilename{LMPIT_uncorrelated}
		\input{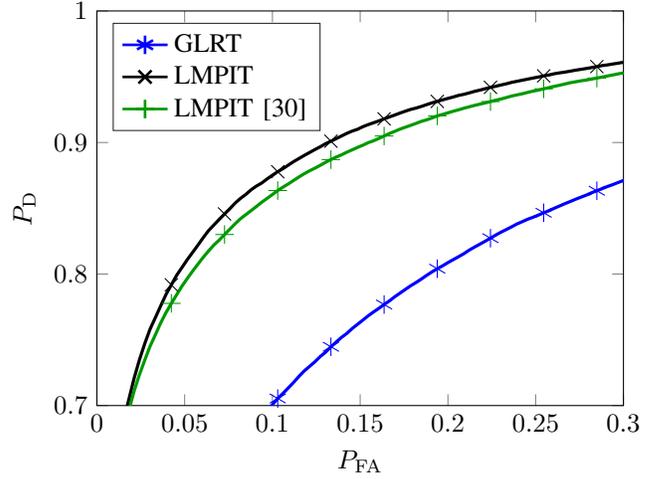}
		\caption{ROC curves for detecting a cyclostationary signal in uncorrelated noise.}
		\label{fig:uncorrelated}
\end{figure}

The performance is illustrated in \autoref{fig:uncorrelated} by means of an ROC curve, which depicts the probability of detection $P_\mathrm{D}$ and the probability of false alarm $P_\mathrm{FA}$. Interestingly, both LMPITs outperform the GLRT, even though the LMPIT from \cite{Ramirez2015b} does not exploit the additional information about spatial uncorrelatedness. The two LMPITs perform very similarly, and we do not gain much by taking into account spatial uncorrelatedness. We already observed something similar in a simulation with spatially uncorrelated noise in \cite{Pries2016}, where we compared the GLRT for the case of uncorrelated noise with the GLRT for correlated noise. This is due to the fact that the number of unknown parameters is not reduced as much for the case of spatially uncorrelated noise.

\subsection{Temporally White Noise}
For the case of temporally white noise, we use the LMPIT-inspired tests proposed in \autoref{sec:LMPIT_inspired}. Further, there is also the LMPIT in \cite{Ramirez2015b} and the GLRT presented in \autoref{sec:GLRT}, which can be used for this scenario. We then have the following list of test statistics:
\begin{align}
\text{\ding{172}} \enskip& \left\lVert\left(\diag{L}{\bhat{S}}\right)^{-1/2}\bhat{S}_1\left(\diag{L}{\bhat{S}}\right)^{-1/2}\right\rVert^2_F\label{test_1}\\
\text{\ding{173}} \enskip& \prod\limits_{j=0}^{N-1}\det\left(\bhat{C}_j\right)\label{test_2}\\
\text{\ding{174}} \enskip& \sum\limits_{j=0}^{N-1}\|\bhat{C}_j\|^2_F\label{test_3}\\
\text{\ding{175}} \enskip& \|\bhat{C}_\text{av}\|^2_F\label{test_4}
\end{align}
The last three test statistics are specific to the scenario of detecting cyclostationarity in white noise, while the first expression (the LMPIT from \cite{Ramirez2015b}) covers the more general case of a cyclostationary signal in temporally colored noise. But because none of the last three tests is optimal, there is no guarantee that they perform better, even though they use additional information.

\subsubsection{ROC curves}
\begin{figure}[!t]
	\centering
	\tikzsetnextfilename{LMPIT_ROC_12}
	\input{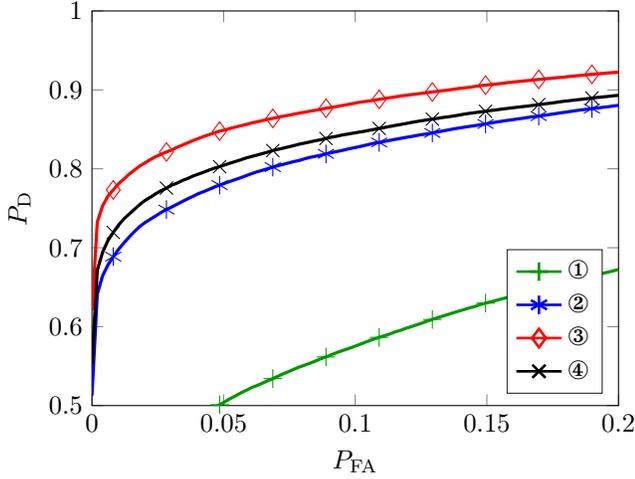}
	\caption{ROC of the proposed tests in \eqref{test_1}--\eqref{test_4}, using $N=12$.}
	\label{fig:ROC_12}
\end{figure}

\begin{figure}[!t]
	\centering
	\tikzsetnextfilename{LMPIT_ROC_64}
	\input{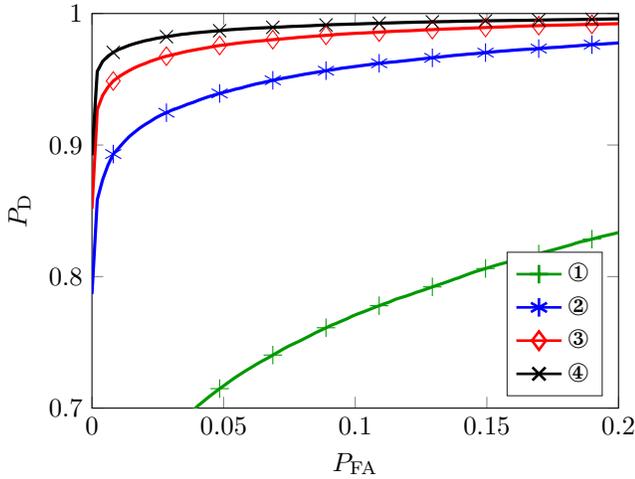}
	\caption{ROC of the proposed tests in \eqref{test_1}--\eqref{test_4}, using $N=64$.}
	\label{fig:ROC_64}
\end{figure}

Figures \ref{fig:ROC_12} and \ref{fig:ROC_64} show ROC curves for simulations with the parameters $\mathrm{SNR}=\SI{-12}{dB}$, $P=3$, $L=3$, $M=20$, as well as white and correlated noise $\mathbf{w}[n]$. The difference between the simulations shown in Figures \ref{fig:ROC_12} and \ref{fig:ROC_64} is the parameter $N$, which is $12$ and $64$, respectively. First of all, the result reveals that our two proposed statistics \ding{174} and \ding{175} perform better than the white-noise GLRT \ding{173} and the colored-noise LMPIT \ding{172}, independently of $N$. Increasing $N$ to $64$ reveals an interesting change of the ordering between \ding{174} and \ding{175}. As illustrated in \autoref{fig:ROC_12}, the statistic \ding{174} leads to a better test for a low $N$, and conversely, the test \ding{175} performs better if $N$ is large (\autoref{fig:ROC_64}).

This phenomenon was also observed for other values of $N$ and $P$: If $N$ or $P$ are large, then the test based on \ding{175} is the best and \ding{174} performs somewhat worse. If $N$ and $P$ are small, then \ding{174} performs best and \ding{175} looses performance.

\subsubsection{Distribution under the null hypothesis}
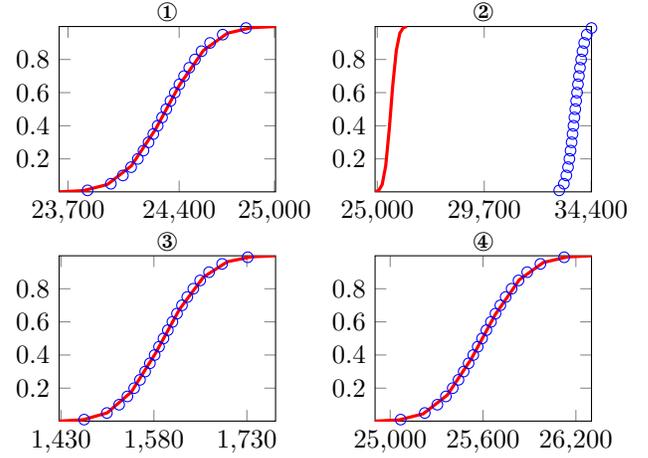
\begin{figure}[!t]
	\centering
	\tikzsetnextfilename{distribution_H0}
%
%
\begin{tikzpicture}

\begin{axis}[%
width=0.411\smallfigurewidth,
height=0.419\smallfigureheight,
at={(0\smallfigurewidth,0.581\smallfigureheight)},
scale only axis,
xmin=23644.1610507504,
xmax=25007.2381984359,
xtick={23700, 24400, 25000},
ymin=0.000999999999999993,
ymax=0.999,
axis background/.style={fill=white},
title style={font=\bfseries},
title={\ding{172}},
scaled x ticks = false,
x tick label style={/pgf/number format/.cd, fixed,precision=3},
title style={yshift=-1.7ex}
]
\addplot [color=blue, draw=none, mark size=2pt, mark=o, mark options={solid, blue}, forget plot]
  table[row sep=crcr]{%
23822.8489315721	0.01\\
23969.8054354455	0.05\\
24045.5245239488	0.1\\
24097.4244395764	0.15\\
24139.7443889949	0.2\\
24175.2224006676	0.25\\
24207.3068469981	0.3\\
24237.0856112256	0.35\\
24264.6573969564	0.4\\
24292.3152147823	0.45\\
24318.9447912792	0.5\\
24345.3878783763	0.55\\
24372.6715797144	0.6\\
24401.5480858929	0.65\\
24430.8450912558	0.7\\
24463.4676780264	0.75\\
24499.1371252273	0.8\\
24540.811946736	0.85\\
24594.5879054121	0.9\\
24674.4132637916	0.95\\
24821.1223285141	0.99\\
};
\addplot [color=red, line width=1.2pt, forget plot]
  table[row sep=crcr]{%
23644.1610507504	0.000999999999999993\\
23795.6140671599	0.00837924979844281\\
23947.0670835694	0.0448807553166099\\
24098.5200999789	0.157621901622442\\
24249.9731163884	0.376456956305295\\
24401.4261327979	0.644984803919471\\
24552.8791492074	0.854423484681685\\
24704.3321656169	0.958768619402711\\
24855.7851820264	0.992127519254144\\
25007.2381984359	0.999\\
};
\end{axis}

\begin{axis}[%
width=0.411\smallfigurewidth,
height=0.419\smallfigureheight,
at={(0.6\smallfigurewidth,0.581\smallfigureheight)},
scale only axis,
xmin=24902.5106385948,
xmax=34413.6980617729,
xtick={25000, 29700, 34400},
ymin=0.00100000000000003,
ymax=0.999,
axis background/.style={fill=white},
title style={font=\bfseries},
title={\ding{173}},
scaled x ticks = false,
x tick label style={/pgf/number format/.cd, fixed,precision=3},
title style={yshift=-1.7ex}
]
\addplot [color=blue, draw=none, mark size=2pt, mark=o, mark options={solid, blue}, forget plot]
  table[row sep=crcr]{%
32995.045096831	0.01\\
33200.8224048761	0.05\\
33308.8100939852	0.1\\
33383.7477389918	0.15\\
33443.0016115082	0.2\\
33493.2568304105	0.25\\
33539.5631055283	0.3\\
33581.7127150541	0.35\\
33622.6661487335	0.4\\
33662.4824376123	0.45\\
33701.3506674209	0.5\\
33740.499398619	0.55\\
33779.4387274462	0.6\\
33819.5894268437	0.65\\
33862.2102226155	0.7\\
33908.0605905481	0.75\\
33959.1512617941	0.8\\
34018.4602345542	0.85\\
34093.0501973937	0.9\\
34205.9994152666	0.95\\
34413.6980617729	0.99\\
};
\addplot [color=red, line width=1.2pt, forget plot]
  table[row sep=crcr]{%
24902.5106385948	0.00100000000000003\\
25057.8859695464	0.00837251255231716\\
25213.2613004981	0.0448322675533716\\
25368.6366314498	0.15746423046119\\
25524.0119624015	0.376182823254931\\
25679.3872933531	0.644717839802024\\
25834.7626243048	0.854277016398506\\
25990.1379552565	0.958724899923034\\
26145.5132862082	0.992121449561541\\
26300.8886171598	0.999\\
};
\end{axis}

\begin{axis}[%
width=0.411\smallfigurewidth,
height=0.419\smallfigureheight,
at={(0\smallfigurewidth,0\smallfigureheight)},
scale only axis,
xmin=1427.0918834838,
xmax=1776.30548921305,
xtick={1430, 1580, 1730},
ymin=0.001,
ymax=0.999,
axis background/.style={fill=white},
title style={font=\bfseries},
title={\ding{174}},
scaled x ticks = false,
x tick label style={/pgf/number format/.cd, fixed,precision=3},
title style={yshift=-1.7ex}
]
\addplot [color=blue, draw=none, mark size=2pt, mark=o, mark options={solid, blue}, forget plot]
  table[row sep=crcr]{%
1467.43308947514	0.01\\
1504.13898069353	0.05\\
1523.88154829695	0.1\\
1537.26238032681	0.15\\
1548.12809727335	0.2\\
1557.52831539801	0.25\\
1565.85528021045	0.3\\
1573.52146830371	0.35\\
1581.17565585695	0.4\\
1588.26061456721	0.45\\
1595.40718437235	0.5\\
1602.46166545438	0.55\\
1609.83157786221	0.6\\
1617.20327564011	0.65\\
1625.26975289021	0.7\\
1633.83050537184	0.75\\
1643.60360488622	0.8\\
1654.62777186193	0.85\\
1668.95549912218	0.9\\
1690.06131770178	0.95\\
1731.33679552466	0.99\\
};
\addplot [color=red, line width=1.2pt, forget plot]
  table[row sep=crcr]{%
1427.0918834838	0.001\\
1465.8933952315	0.00921877503664084\\
1504.69490697919	0.050907829450181\\
1543.49641872688	0.176739845638615\\
1582.29793047458	0.40849704139464\\
1621.09944222227	0.674989288993008\\
1659.90095396997	0.870349359664738\\
1698.70246571766	0.96343041694755\\
1737.50397746535	0.992775352700845\\
1776.30548921305	0.999\\
};
\end{axis}

\begin{axis}[%
width=0.411\smallfigurewidth,
height=0.419\smallfigureheight,
at={(0.6\smallfigurewidth,0\smallfigureheight)},
scale only axis,
xmin=24902.5106385948,
xmax=26300.8886171598,
xtick={25000, 25600, 26200},
ymin=0.00100000000000003,
ymax=0.999,
axis background/.style={fill=white},
title style={font=\bfseries},
title={\ding{175}},
scaled x ticks = false,
x tick label style={/pgf/number format/.cd, fixed,precision=3},
title style={yshift=-1.7ex}
]
\addplot [color=blue, draw=none, mark size=2pt, mark=o, mark options={solid, blue}, forget plot]
  table[row sep=crcr]{%
25067.6908001446	0.01\\
25223.6001973231	0.05\\
25304.9929987161	0.1\\
25360.2358532056	0.15\\
25404.2066378749	0.2\\
25441.2891969244	0.25\\
25474.9424617502	0.3\\
25506.8393955944	0.35\\
25536.9214650505	0.4\\
25565.7342504601	0.45\\
25593.8107256171	0.5\\
25621.5872950234	0.55\\
25651.25226161	0.6\\
25680.4661802478	0.65\\
25711.7791897561	0.7\\
25746.3483250993	0.75\\
25784.5744384132	0.8\\
25829.2723569124	0.85\\
25885.3738828484	0.9\\
25970.9825053212	0.95\\
26124.4545196864	0.99\\
};
\addplot [color=red, line width=1.2pt, forget plot]
  table[row sep=crcr]{%
24902.5106385948	0.00100000000000003\\
25057.8859695464	0.00837251255231716\\
25213.2613004981	0.0448322675533716\\
25368.6366314498	0.15746423046119\\
25524.0119624015	0.376182823254931\\
25679.3872933531	0.644717839802024\\
25834.7626243048	0.854277016398506\\
25990.1379552565	0.958724899923034\\
26145.5132862082	0.992121449561541\\
26300.8886171598	0.999\\
};
\end{axis}
\end{tikzpicture}%
	\caption{Cumulative distribution functions (CDFs) of the test statistics \ding{172}-\ding{175} under the null hypothesis. Expected CDFs according to the $\chi^2$-approximations (red lines) and the estimated CDFs of the normalized test statistics \ding{172}-\ding{175} for simulations (blue circles).}
	\label{fig:distribution_H0}
\end{figure}

As shown in previous sections, the GLR as well as the LMPIT-inspired statistics are approximately $\chi^2$-distributed under the null hypothesis. In \autoref{fig:distribution_H0}, we show the expected cumulative distribution functions (CDFs) as well as the estimated CDFs from simulation results with temporally white and spatially uncorrelated noise and $M=64$.
It can be observed that all terms that depend on a Frobenius norm are approximated very well by the corresponding $\chi^2$ distribution, while the GLR-statistic \ding{173} is quite far away from the expected result.
A similar result for small $M$ was also observed in \cite{Ramirez2015b}, where it was concluded that the Frobenius norm converges much faster to the $\chi^2$ distribution than the log-det.
\subsubsection{Robustness against model misspecification}
In another simulation we tested the robustness of the proposed LMPIT-inspired detectors against violation of the white noise assumption. Instead of white noise, we used noise with increasing temporal correlation. This was achieved by passing white noise through an FIR-filter with impulse response $g[n]=\exp(-\frac{n}{\sigma})$, where $\sigma$ controls the degree of temporal correlation: White noise is obtained for $\sigma\rightarrow 0$ and an increasing $\sigma$ introduces an increasing level of temporal correlation.
\begin{figure}[!t]%
	\centering%
	\tikzsetnextfilename{noise_color_PD}%
%
%
\begin{tikzpicture}

\begin{axis}[%
width=\smallfigurewidth,
height=\smallfigureheight,
at={(0\smallfigurewidth,0\smallfigureheight)},
scale only axis,
xmin=0,
xmax=20,
xlabel={$\sigma$},
ymin=0,
ymax=0.9,
ylabel={$P_{\mathrm{D}}$},
axis background/.style={fill=white},
legend style={legend cell align=left,align=left,draw=white!15!black},
scaled x ticks = false,
x tick label style={/pgf/number format/.cd, fixed,precision=3}
]
\addplot [color=black!40!green,solid,line width=1.2pt,mark size=3.5pt,mark=+,mark options={solid}]
  table[row sep=crcr]{%
0	0.1787\\
2.85714285714286	0.1809\\
5.71428571428571	0.2587\\
8.57142857142857	0.3199\\
11.4285714285714	0.375\\
14.2857142857143	0.4441\\
17.1428571428571	0.4612\\
20	0.5003\\
};
\addlegendentry{\ding{172}};

\addplot [color=blue,solid,line width=1.2pt,mark size=3.5pt,mark=asterisk,mark options={solid}]
  table[row sep=crcr]{%
0	0.6779\\
2.85714285714286	0.7848\\
5.71428571428571	0.5717\\
8.57142857142857	0.4454\\
11.4285714285714	0.3723\\
14.2857142857143	0.3242\\
17.1428571428571	0.2709\\
20	0.2538\\
};
\addlegendentry{\ding{173}};

\addplot [color=red,solid,line width=1.2pt,mark size=3.5pt,mark=diamond,mark options={solid}]
  table[row sep=crcr]{%
0	0.7782\\
2.85714285714286	0.3119\\
5.71428571428571	0.0008\\
8.57142857142857	0.0001\\
11.4285714285714	0.0001\\
14.2857142857143	0.0001\\
17.1428571428571	0.0001\\
20	0.0001\\
};
\addlegendentry{\ding{174}};

\addplot [color=black,solid,line width=1.2pt,mark size=3.5pt,mark=x,mark options={solid}]
  table[row sep=crcr]{%
0	0.8848\\
2.85714285714286	0.8578\\
5.71428571428571	0.6379\\
8.57142857142857	0.4302\\
11.4285714285714	0.2719\\
14.2857142857143	0.175\\
17.1428571428571	0.1224\\
20	0.096\\
};
\addlegendentry{\ding{175}};

\end{axis}
\end{tikzpicture}%
%
	\caption{$P_\mathrm{D}$ at $P_{\mathrm{FA}}=0.01$ for noise with increasing temporal correlation controlled by $\sigma$.}%
	\label{fig:noise_color_PD}%
\end{figure}
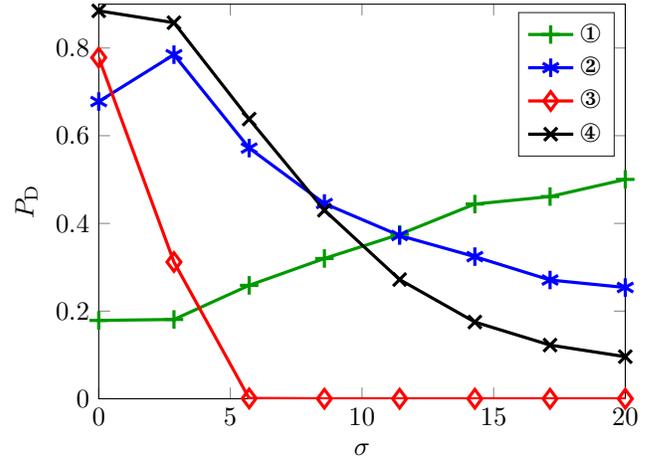%

Figure \ref{fig:noise_color_PD} shows the resulting performance as measured by the probability of detection $P_\mathrm{D}$, at $P_{\mathrm{FA}}=0.01$. In this simulation we further use $\mathrm{SNR}=\SI{-16}{dB}$, $P=4$, $N=64$, $L=3$, $M=20$ and QPSK signals with RRC pulse shaping. As expected, the detectors designed for the white-noise scenario (\ding{173}-\ding{175}) perform well for the case of almost white noise and the general-noise LMPIT \ding{172} performs best when the temporal correlation is large. Interestingly, the tests \ding{173} and \ding{175} are more robust against deviation from white noise, as opposed to the test based on \ding{174}. This robustness also holds when the simulation is performed for a small $N$, for example $N=12$. Thus, the detector \ding{175} should in most cases be preferred over \ding{174}.

\subsubsection{Comparison with state-of-the-art detectors}
Now we compare the performance of the proposed detectors with detectors from \cite{Urriza2013,Lunden2009}. We use OFDM modulation with a QPSK constellation to demonstrate that the results are not specific to single-carrier modulations. Since \cite{Urriza2013,Lunden2009} do not require $M>1$, we simulated the received signals with one long observation for a fair comparison. This long observation was split into multiple segments when using the statistics \ding{172}-\ding{175}. In particular, we use $L=2$ antennas and receive $1024$ symbols of OFDM signals with $16$ subcarriers and a cyclic prefix of $4$, sampled at Nyquist rate, which results in $20$ samples per symbol \cite{Lunden2009}. Thus the cycle period is $P=20$. For the detectors \ding{172}-\ding{175}, we factor $NM=1024$ into $M=64$ segments of length $N=16$. For both \cite{Urriza2013,Lunden2009} we use the first cycle frequency, while the lags are chosen as $\pm 16$ and $16$ for \cite{Lunden2009} and \cite{Urriza2013}, respectively. This choice incorporates prior information about the maximum of the cyclic autocorrelation function for OFDM signals \cite{Oner2007, Lunden2009}. In a practical cognitive radio application, such prior information might not be available. Hence, the comparisons are overly favorable for our competitors.\par
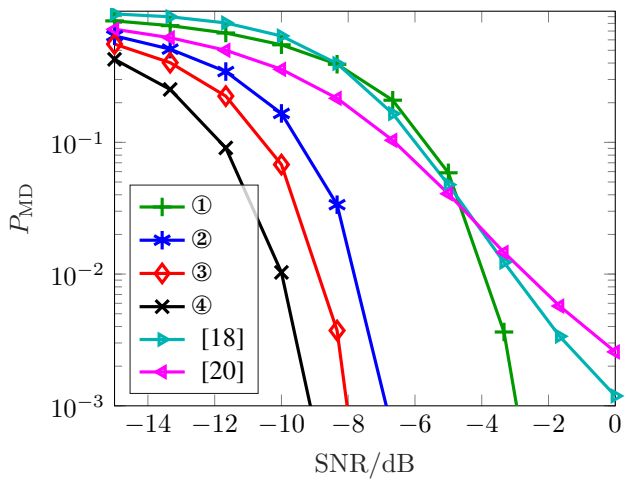
\begin{figure}[!t]
	\centering
	\tikzsetnextfilename{Nyquist_SNR}
%
%
\definecolor{mycolor1}{rgb}{0.00000,0.70000,0.70000}%
\definecolor{mycolor2}{rgb}{1.00000,0.00000,1.00000}%
\begin{tikzpicture}

\begin{axis}[%
width=0.951\smallfigurewidth,
height=\smallfigureheight,
at={(0\smallfigurewidth,0\smallfigureheight)},
scale only axis,
xmin=-15,
xmax=0,
xlabel style={font=\color{white!15!black}},
xlabel={$\mathrm{SNR} / \si{dB}$},
ymode=log,
ymin=1e-03,
ymax=0.99,
yminorticks=true,
ylabel style={font=\color{white!15!black}},
ylabel={$P_{\mathrm{MD}}$},
axis background/.style={fill=white},
legend style={at={(0.03,0.03)}, anchor=south west, legend cell align=left, align=left, draw=white!15!black,fill opacity=0.8,text opacity =1},
scaled x ticks = false,
x tick label style={/pgf/number format/.cd, fixed,precision=3},
title style={yshift=-1.7ex}
]
\addplot [color=black!40!green, line width=1.2pt, mark size=3.5pt, mark=+, mark options={solid, black!40!green}]
  table[row sep=crcr]{%
-15	0.835871666666667\\
-13.3333333333333	0.769521666666667\\
-11.6666666666667	0.676459166666667\\
-10	0.550474166666667\\
-8.33333333333333	0.3916775\\
-6.66666666666667	0.208936666666667\\
-5	0.0587466666666667\\
-3.33333333333333	0.00363500000000005\\
-1.66666666666667	1.24999999999709e-05\\
0	0\\
};
\addlegendentry{$\text{\ding{172}}$}

\addplot [color=blue, line width=1.2pt, mark size=3.5pt, mark=asterisk, mark options={solid, blue}]
  table[row sep=crcr]{%
-15	0.643721666666667\\
-13.3333333333333	0.510075833333333\\
-11.6666666666667	0.34482\\
-10	0.1658875\\
-8.33333333333333	0.0336416666666667\\
-6.66666666666667	0.000631666666666586\\
-5	0\\
-3.33333333333333	0\\
-1.66666666666667	0\\
0	0\\
};
\addlegendentry{$\text{\ding{173}}$}

\addplot [color=red, line width=1.2pt, mark size=3.5pt, mark=diamond, mark options={solid, red}]
  table[row sep=crcr]{%
-15	0.557499166666667\\
-13.3333333333333	0.402736666666667\\
-11.6666666666667	0.2243825\\
-10	0.0678183333333333\\
-8.33333333333333	0.00373083333333335\\
-6.66666666666667	2.50000000001638e-06\\
-5	0\\
-3.33333333333333	0\\
-1.66666666666667	0\\
0	0\\
};
\addlegendentry{$\text{\ding{174}}$}

\addplot [color=black, line width=1.2pt, mark size=3.5pt, mark=x, mark options={solid, black}]
  table[row sep=crcr]{%
-15	0.4268225\\
-13.3333333333333	0.2523\\
-11.6666666666667	0.0907825\\
-10	0.0103033333333333\\
-8.33333333333333	0.000115833333333315\\
-6.66666666666667	0\\
-5	0\\
-3.33333333333333	0\\
-1.66666666666667	0\\
0	0\\
};
\addlegendentry{$\text{\ding{175}}$}

\addplot [color=mycolor1, line width=1.2pt, mark size=2.3pt, mark=triangle, mark options={solid, rotate=270, mycolor1}]
  table[row sep=crcr]{%
-15	0.944001666666667\\
-13.3333333333333	0.898240833333333\\
-11.6666666666667	0.806755\\
-10	0.642105\\
-8.33333333333333	0.396684166666667\\
-6.66666666666667	0.165310833333333\\
-5	0.0477183333333334\\
-3.33333333333333	0.0123116666666666\\
-1.66666666666667	0.0033725\\
0	0.00118833333333335\\
};
\addlegendentry{\cite{Lunden2009}}

\addplot [color=mycolor2, line width=1.2pt, mark size=2.3pt, mark=triangle, mark options={solid, rotate=90, mycolor2}]
  table[row sep=crcr]{%
-15	0.722494166666667\\
-13.3333333333333	0.622800833333333\\
-11.6666666666667	0.4999925\\
-10	0.358625833333333\\
-8.33333333333333	0.216053333333333\\
-6.66666666666667	0.10365\\
-5	0.0407316666666666\\
-3.33333333333333	0.0146116666666667\\
-1.66666666666667	0.00571833333333338\\
0	0.00256666666666661\\
};
\addlegendentry{\cite{Urriza2013}}

\end{axis}
\end{tikzpicture}%
	\caption{Probability of missed detection $P_\mathrm{MD}$ at $P_\mathrm{FA}=0.01$ for varying $\mathrm{SNR}$ using OFDM transmission.}
	\label{fig:Nyquist_SNR}
\end{figure}

The performance of the selected detectors for various $\mathrm{SNR}$s is illustrated in \autoref{fig:Nyquist_SNR}. It can be seen that the test \ding{175} also performs best in this setup. The tests not specific to the white-noise scenario (i.e.\ \ding{172},  \cite{Lunden2009}, and \cite{Urriza2013}), however, perform considerably worse.

\section{Computational complexity}
\label{sec:complexity}
In this section, we estimate the computational complexity of our detectors in terms of floating point operations (FLOPs). To approximate the complexity, we focus on the most time-consuming parts of the algorithm, which are equations \eqref{eq:transformation}, \eqref{eq:S_full}, \eqref{eq:C_hat}, and then the tests themselves. For matrix operations, we use the FLOP estimates from \cite{Hunger2007}.

Equation \eqref{eq:transformation} is most efficiently implemented by an FFT. Since we need $LM$ FFTs of length $NP$, this requires approximately $5LMNP\log_2(NP)$ FLOPS \cite{VanLoan1992}. Next we only need to compute the diagonal blocks of \eqref{eq:S_full}, which costs approximately $MN(LP)^2$ FLOPS. Finally, we can compute the inverse in \eqref{eq:C_hat} by inverting the $L\times L$ diagonal blocks. In terms of computational complexity this is negligible compared to the rest of the matrix multiplication in \eqref{eq:C_hat}, which in turn can be optimized by exploiting the block-diagonal structure of the involved matrices. Thus this operation takes approximately $2NL^3P^2$ FLOPS for the case of spatially correlated noise. If we use a detector for the case of spatially uncorrelated noise, this is reduced to $NL^2P^2$ FLOPS. On top of this, the detectors need to be computed. The LMPIT or LMPIT-inspired tests compute the Frobenius norm, which is only of linear complexity in the matrix size. The determinant for the GLRTs has a bigger impact with approximately $\frac{1}{3}N(LP)^3$ FLOPS.

To summarize, for a large sensing duration (i.e.~$N$), the FFT is the computational bottleneck. Since competing detectors typically also use FFT-based statistics, the asymptotic complexity in $N$ is similar to our detectors. Our detectors further benefit from the fact that they only require standard matrix operations and the FFT. These operations exist in many standard math libraries and are often optimized with respect to other parameters such as memory and cache. Our detectors can further benefit from parallelization.

\section{Noise Characterization}%
\label{sec:characterization}
So far we have assumed to \emph{know} whether the noise has a particular temporal or spatial structure. If it is not known a priori whether such a structure is present, and consequently which detector is appropriate, we must first detect the noise structure. To this end, we will assume to have available samples of noise only. \footnote{An alternative approach, which does not need noise-only samples, would be a \emph{multiple hypothesis} test. Then the different hypotheses correspond to a signal that is either cyclostationary, WSS with arbitrary spatio-temporal correlation, or WSS without spatial and/or temporal correlation. As a multiple hypothesis test is out of the scope of this paper, we do not follow this alternative approach.}%

Testing whether or not a process is temporally white has been treated in \cite{Box1970,Ljung1978} and extensions for multivariate processes have been published in \cite{Chitturi1974,Hosking1980,Mahdi2012}. Tests for spatial (un)-correlatedness of random vectors were derived in \cite{Wilks1935,Leshem2001,Ramirez2010}. Since it is possible to derive asymptotic tests in the framework of this paper, we now present GLRTs to determine if the noise is temporally white/colored or spatially uncorrelated/correlated. To keep the same notation as before, we assume to have $NP$ samples of the noise process $\mathbf{x}[n]=\mathbf{w}[n]$. As in \autoref{sec:problem}, we collect all samples in the vector $\mathbf{y}$ and transform it to $\mathbf{z}$. Given multiple realizations of $\mathbf{y}$, we test whether or not some temporal or spatial structure is present.
\subsection{Testing the Temporal Structure}
Here we consider the hypotheses
\begin{equation*}
\begin{split}
\mathcal{H}_1&: \text{$\mathbf{x}[n]$ is temporally colored,} \\
\mathcal{H}_0&: \text{$\mathbf{x}[n]$ is temporally white.}
\end{split}
\label{eq:whiteHypothesis}
\end{equation*}
With the Gaussian assumption, these hypotheses are asymptotically equivalent to
\begin{equation*}
\begin{split}
\mathcal{H}_1&: \mathbf{z} \sim \mathcal{CN}\left(\mathbf{0},\mathbf{S}_1\right)\\
\mathcal{H}_0&: \mathbf{z} \sim \mathcal{CN}\left(\mathbf{0},\mathbf{I}_{NP}\otimes\mathbf{S}_0\right),
\end{split}
\label{eq:whiteHypothesis2}
\end{equation*}
where $\mathbf{S}_1$ is a block-diagonal matrix with blocks of size $L\times L$, and $\mathbf{S}_0$ is an $L \times L$ matrix. Thus we essentially test whether or not the diagonal blocks of the sample covariance matrix are identical. Since we do not know these blocks, we have to estimate them, which leads to a GLRT. The ML-estimates are listed in \autoref{tab:S0_hat}, and using them we find an expression for the log-GLR:
\begin{equation}
\sum_{k=0}^{NP-1}\log\det\left(\bhat{S}^{(k,k)}\right)-NP\log\det\left(\frac{1}{NP}\sum_{k=0}^{NP-1}\bhat{S}^{(k,k)}\right).\label{eq:GLR_color}
\end{equation}
The log-GLRT is obtained by comparing \eqref{eq:GLR_color} with a threshold $\eta$. If it is smaller than $\eta$, the (white noise) null hypothesis is rejected.

\subsection{Testing the Spatial Structure}
The hypotheses for testing the spatial structure are
\begin{equation*}
\begin{split}
\mathcal{H}_1&: \text{$\mathbf{x}[n]$ is spatially correlated,}\\
\mathcal{H}_0&: \text{$\mathbf{x}[n]$ is spatially uncorrelated.}
\end{split}
\label{eq:uncorrelatedHypothesis}
\end{equation*}
Asymptotically, this is equivalent to
\begin{equation*}
\begin{split}
\mathcal{H}_1&: \mathbf{z} \sim \mathcal{CN}\left(\mathbf{0},\mathbf{S}_1\right)\\
\mathcal{H}_0&: \mathbf{z} \sim \mathcal{CN}\left(\mathbf{0},\mathbf{S}_0\right),
\end{split}
\label{eq:uncorrelatedHypothesis2}
\end{equation*}
where $\mathbf{S}_1$ is a block-diagonal matrix with blocks of size $L \times L$ and $\mathbf{S}_0$ is a diagonal matrix. Thus the present test is a special case of the test between two block-diagonal matrices from \cite{Ramirez2015b}, and we can specialize the test to the block sizes $L$ and $1$. Then the log-GLR can be written as
\begin{equation}
\sum_{k=0}^{NP-1}\log\det\left(\bhat{S}^{(k,k)}\right)-\log\det\left(\diag{}{\bhat{S}}\right)
\label{eq:GLR_correlation}
\end{equation}
and the (uncorrelated noise) null hypothesis is rejected for small values.

%
\ifCLASSOPTIONdraftcls%
\else%
\begin{figure*}[!tb]%
	\hrulefill%
	\setcounter{MYtempeqncnt2}{\value{equation}}
	\setcounter{equation}{65} 
	\input{sum_permutations.tex}%
	\setcounter{equation}{\value{MYtempeqncnt2}}
	\hrulefill%
\end{figure*}%
\fi%
\section{Conclusions}
We have presented tests for the detection of a cyclostationary signal with known cycle period in noise with known statistical properties. In the case of temporally colored and spatially uncorrelated noise, it was possible to find an LMPIT, which computes the Frobenius norm of a sample coherence matrix. Thus we obtained the same result as in the case of spatially correlated noise, where the LMPIT and the GLRT are, respectively, the Frobenius norm and the determinant of another sample coherence matrix. As shown in simulations, the performance gain compared to the LMPIT for noise with arbitrary spatial correlation is small.

The case of white noise is quite different. Here the LMPIT does not exist, as the likelihood ratio of the maximal invariant statistics depends on unknown quantities. Instead, we proposed two LMPIT-inspired tests. These tests are suboptimal, but it was shown in simulations that these detectors can outperform other tests for a variety of scenarios. This includes the case of communications signals where the distribution under the alternative is not complex normal and the case when only one realization is available. The thresholds for the tests that depend on a Frobenius norm can be chosen using a $\chi^2$ distribution.

Finally, we considered the case where a-priori information about the noise structure is not available. If noise-only samples are available, we have also proposed tests to infer the noise structure. This enables the utilization of the appropriate detector for the subsequent signal detection task. %

MATLAB code for our detectors is available at \url{https://github.com/SSTGroup/Cyclostationary-Signal-Processing}.

\appendices
\section{Proof of Lemma~\ref{lemma1}}
\label{ap:lemma1}
The proof follows along the lines of the derivation of the LMPIT in \cite{Ramirez2015b}. %
First we note that the determinants of $\mathbf{S}_0$ and $\mathbf{S}_1$ in (\ref{eq:wijsman}) are constant with respect to the observations and thus they are irrelevant for the test statistic. Next we see that $\tilde{\mathbf{G}}^H\mathbf{S}_1^{-1}\tilde{\mathbf{G}}$ as well as $\tilde{\mathbf{G}}^H\mathbf{S}_0^{-1}\tilde{\mathbf{G}}$ are block-diagonal with block-size $LP$. For this reason, the traces only depend on the diagonal $LP\times LP$ blocks of $\bhat{S}$ and thus we can replace $\bhat{S}$ by $\bhat{S}_1=\diag{LP}{\bhat{S}}$ without changing the outcome. Now we introduce the change of variables 
\begin{equation}
\mathbf{G}\rightarrow\mathbf{G}\left(\frac{1}{NP}\sum\limits_{j=0}^{N-1}\sum\limits_{k=0}^{P-1}\bhat{S}_j^{(k,k)}\right)^{-1/2}
\end{equation}
in the nominator and the denominator. Both steps combined cause the normalization $\Shat_1\rightarrow\hat{\mathbf{C}}$ with the coherence matrix $\bhat{C} =  \bhat{S}_0^{-1/2}\bhat{S}_1\bhat{S}_0^{-1/2}$ and
\begin{equation}
\bhat{S}_0 = \mathbf{I}_{NP}\otimes\frac{1}{NP}\sum\limits_{j=0}^{N-1}\sum\limits_{k=0}^{P-1}\bhat{S}_j^{(k,k)}.
\end{equation}
Applying the transformation 
\begin{equation}
\mathbf{G}\rightarrow\left(\frac{1}{NP}\sum\limits_{j=0}^{N-1}\sum\limits_{k=0}^{P-1}\bhat{S}_j^{(k,k)}\right)^{+1/2}\mathbf{G},
\end{equation}
we see that the trace in the denominator is constant:
\begin{equation}
\tr{\btilde{G}\bhat{C}\btilde{G}^H} = NP\tr{\mathbf{GG}^H}.
\end{equation}
Thus the denominator does not depend on data and can be discarded. Applying another transformation $\mathbf{G}\rightarrow\bar{\mathbf{S}}_1^{-1/2}\mathbf{G}$ causes $\mathbf{S}_1^{-1} \rightarrow\tilde{\mathbf{S}}_1$. Finally, we rewrite and simplify the integral in terms of $\alpha$, $\beta(\mathbf{G})$ and $\mathbf{W}$. This concludes the proof.
\section{Proof of Lemma~\ref{lemma_linear}}
\label{ap:lemma2}
We define
\begin{align}
\bm{\Psi} = \sum\limits_{\mathbb{P}}\int\limits_{\mathbb{Q}}\int\limits_{\mathbb{G}}\beta(\mathbf{G})\mathbf{W}\id\mathbf{G}\id\mathbf{Q},
\end{align}
and note that \eqref{eq:LMPIT_approx_linear} can be expressed as $\cramped{\tr{\bm{\Psi}\bhat{C}}}$. Since $\mathbf{W}$ is block-diagonal with blocks of size $LP\times LP$, so is $\bm{\Psi}$. The permutation matrix $\mathbf{P}$ permutes these blocks and by summing over all possible permutations, the blocks become identical. This can be expressed as
\begin{equation}
\bm{\Psi} = \mathbf{I}_N\otimes\bm{\Phi},
\end{equation}
where $\bm{\Phi}$ is an $LP\times LP$ matrix. Now we have a similar problem as in \cite{Ramirez2013} and following the proof therein, it can be shown that $\bm{\Phi}$ is a diagonal matrix with identical elements. Therefore we can simplify \eqref{eq:LMPIT_approx_linear}:
\begin{align}
\tr{\bm{\Psi}\bhat{C}} \propto \tr{\bhat{C}} = NLP,
\end{align}
because of the way $\bhat{C}$ is normalized.%
\section{Proof of Lemma~\ref{lemma_quadratic}}
\label{ap:lemma3}
Since $\mathbf{W}$ and $\bhat{C}$ are block-diagonal, we can first express the trace in terms of their diagonal blocks of size $LP\times LP$:
\begin{equation}
\begin{multlined}
\trsq{\mathbf{W}\bhat{C}}=\sum\limits_{j=0}^{N-1}\trsq{\mathbf{W}_j\bhat{C}_j}\\
+\sum\limits_{j=0}^{N-1}\sum\limits_{\substack{i=0\\i\neq j}}^{N-1}\tr{\mathbf{W}_j\bhat{C}_j}\tr{\mathbf{W}_i\bhat{C}_i}.
\label{eq:trace_squared}
\end{multlined}
\end{equation}
Now we take care of the permutations. Note that the permutation matrix $\mathbf{P}$ in $\mathbf{W}$ permutes the set of blocks $\mathbf{W}_j$. With this in mind we sum over all permutations of \eqref{eq:trace_squared}. Using induction it is possible to show that the result can be written as %
\ifCLASSOPTIONdraftcls%
\begin{align}
	\sum_\mathbb{P}\trsq{\mathbf{W}\bhat{C}} &= (N-2)!\cdot\left( N\sum\limits_{i=0}^{N-1}\sum\limits_{j=0}^{N-1}\trsq{\mathbf{W}_i\hat{\mathbf{C}}_j}
	+ N^4\trsq{\mathbf{W}_\text{av}\hat{\mathbf{C}}_\text{av}}\right)\nonumber\\
	&- (N-2)!\cdot N^2\left(\sum\limits_{j=0}^{N-1}\trsq{\mathbf{W}_\text{av}\hat{\mathbf{C}}_j}+
	\sum\limits_{j=0}^{N-1}\trsq{\mathbf{W}_j\hat{\mathbf{C}}_\text{av}}\right)%
	\label{eq:permutations}%
\end{align}
\else%
\stepcounter{equation}%
stated in \eqref{eq:permutations}.
\fi%
Here we introduced the matrix
\begin{equation}
	\hat{\mathbf{W}}_{\mathrm{av}} = \frac{1}{N}\sum\limits_{j=0}^{N-1}\bhat{W}_j,
\end{equation}
and $\hat{\mathbf{C}}_{\mathrm{av}}$ was defined in \eqref{eq:C_av}.
Plugging this result back into \eqref{eq:quadratic_term}, the integrals are now expressed in terms of the blocks $\mathbf{W}_j$ and $\bhat{C}_j$. Since 
\begin{equation}
\mathbf{W}_j = \left(\mathbf{Q}\otimes\mathbf{G}\right)^H\left(\btilde{S}_{1,j}-\mathbf{I}\right)\left(\mathbf{Q}\otimes\mathbf{G}\right),
\end{equation}
the problem at first looks very similar to the one in \cite{Ramirez2013}. In fact, the problems are identical for the case of $N=1$. However, for $N>1$, we have multiple terms in \eqref{eq:trace_squared}. Moreover, the normalization of $\mathbf{W}_j$ and $\bhat{C}_j$ as defined for this problem is different compared to the counterparts in \cite{Ramirez2013}, and thus requires a different solution.

The next step is to express the squared traces in \eqref{eq:permutations} in terms of the $L\times L$ sub-blocks of the $LP\times LP$ blocks $\mathbf{W}_j$ and $\bhat{C}_j$. For the present problem, we can use the invariances in the same way as in \cite[Lemmas~5-7]{Ramirez2013}, but due to the different normalization, fewer terms are constant. Most importantly, the sum of the diagonal sub-blocks is \emph{not} normalized, i.e.\ generally, $\sum_{k=0}^{P-1}\bhat{C}_j^{(k,k)} = P\mathbf{I}$ only if $N=1$. Accounting for this difference, the rest of the proof follows along the lines of \cite[Lemmas~5-7]{Ramirez2013}. Then the quadratic term \eqref{eq:quadratic_term} in the Taylor series expansion can be written as
\begin{align}
\mathscr{L} &\propto c_1\sum\limits_{j=0}^{N-1}\|\hat{\mathbf{C}}_j\|_F^2 + c_2 P\sum\limits_{j=0}^{N-1}\|\hat{\bar{\mathbf{C}}}_j\|_F^2+c_3 N\|\hat{\mathbf{C}}_{\mathrm{av}}\|_F^2,
\end{align}
with unknown $c_1$, $c_2$, and $c_3$. After defining $\lambda = \frac{c_2}{c_1}$ and $\mu = \frac{c_3}{c_1}$, this can be rewritten as stated in \eqref{eq:LMPIT}.
%


\bibliography{bibliography}

\end{document}